\documentclass[reprint,aps,pra,floatfix]{revtex4-1}
\usepackage[english]{babel}
\usepackage{amsmath,amssymb,bm,bbm}
\usepackage{graphicx,xcolor,xspace}
\usepackage{hyperref}
\xspaceaddexceptions{]\}+}

\newcommand{\Z}{\mathbb{Z}}

\renewcommand{\O}{\mathcal{O}}

\newcommand{\TT}{\mathcal{T}}

\newcommand{\up}{\uparrow}
\newcommand{\dn}{\downarrow}
\newcommand{\ket}[1]{{\lvert#1\rangle}}

\newcommand{\lrangle}[1]{\langle#1\rangle}

\newcommand{\xFM}{\ensuremath{x\!\operatorname{FM}}\xspace}
\newcommand{\yAFM}{\ensuremath{y\!\operatorname{AFM}}\xspace}
\newcommand{\zFM}{\ensuremath{z\!\operatorname{FM}}\xspace}
\newcommand{\VBS}{\ensuremath{\operatorname{VBS}}\xspace}
\newcommand{\VBSI}{\ensuremath{\operatorname{VBS-I}}\xspace}
\newcommand{\VBSII}{\ensuremath{\operatorname{VBS-II}}\xspace}

\newcommand{\dd}{\mathrm{d}}
\newcommand{\ii}{\mathrm{i}\,}
\newcommand{\phigood}{\tilde{\phi}}
\newcommand{\thtgood}{\tilde{\theta}}

\begin{document}

\title{Deconfined quantum critical point in one dimension}

\date{\today}

\author{Brenden Roberts}
\email{broberts@caltech.edu}
\author{Shenghan Jiang}
\email{jiangsh@caltech.edu}
\author{Olexei I.~Motrunich}
\email{motrunch@caltech.edu}
\affiliation{Institute for Quantum Information and Matter,\\California Institute of Technology, Pasadena, CA 91125}

\begin{abstract}
We perform a numerical study of a spin-1/2 model with $\Z_2 \times \Z_2$ symmetry in one dimension which demonstrates an interesting similarity to the physics of two-dimensional deconfined quantum critical points (DQCP).
Specifically, we investigate the quantum phase transition between Ising ferromagnetic and valence bond solid (VBS) symmetry-breaking phases.
Working directly in the thermodynamic limit using uniform matrix product states, we find evidence for a direct continuous phase transition that lies outside of the Landau--Ginzburg--Wilson paradigm.
In our model, the continuous transition is found everywhere on the phase boundary.
We find that the magnetic and VBS correlations show very close power law exponents, which is expected from the self-duality of the parton description of this DQCP.
Critical exponents vary continuously along the phase boundary in a manner consistent with the predictions of the field theory for this transition.
We also find a regime where the phase boundary splits, as suggested by the theory, introducing an intermediate phase of coexisting ferromagnetic and VBS order parameters.
Interestingly, we discover a transition involving this coexistence phase which is similar to the DQCP, being also disallowed by Landau--Ginzburg--Wilson symmetry-breaking theory.
\end{abstract}

\maketitle

\section{Introduction}
\label{sec:intro}

The {\it deconfined quantum critical point} (DQCP) is a fascinating proposal by \textcite{DQCP_science, DQCP_prb} whereby a quantum many-body system undergoes a generic second-order transition between phases with incompatible order parameters, which is prohibited in the Landau--Ginzburg--Wilson (LGW) symmetry-breaking paradigm.
The original work predicted such a transition from a Neel phase to a Valence Bond Solid phase in a spin-1/2 system on a two-dimensional (2d) square lattice, and extensive numerical studies have provided evidence for a continuous (or weakly first order) transition~\cite{Sandvik2007, MelkoKaul2008, LuoSandvikKawashima2009, BanerjeeDamleAlet2010, Sandvik2010, HaradaSuzukiOkuboMatsuoLuoWatanabeTodoKawashima2013, JiangNyfelerChandrasekharanWises2008, ChenHuangDengKuklovProkofevSvistunov2013, NahumChalkerSernaOrtunoSomoza2015, NahumSernaChalkerOrtunoSomoza2015II, MotrunichVishwanath2008, KuklovMatsumotoProkofevSvistunovTroyer2008, Bartosch2013, CharrierAletPujol2008, ChenGukelbergerTrebstFabienBalents2009, CharrierAlet2010, SreejithPowell2015, ShaoGuoSandvik2016}.
This DQCP was revisited recently in light of improvements in the understanding of the interplay between symmetries and dualities~\cite{KarchTong2016, SeibergSenthilWangWitten2016, WangNahumMetlitskiXuSenthil2017, MrossAliceaMotrunich2017}, stimulating additional numerical studies~\cite{QinHeYouLuSenSandvikXuMeng2017, GeraedtsMotrunich2017, SernaNahum2018}, although some questions about the transition remain (for a very recent review, see Ref.~\onlinecite{SenthilSonWangXu2018}).

In this paper, we numerically study a simpler version of DQCP realized in one dimension (1d), following the recent theoretical proposal~\cite{jiang19} of a continuous quantum phase transition in a particular 1d model having Ising-type $\Z_2 \times \Z_2$ symmetry as well as translation symmetry.
The transition is between an Ising ferromagnet and a Valence Bond Solid (VBS); as was the case for the 2d DQCP, the phases on either side break different symmetries and a continuous phase transition is disallowed in Landau--Ginzburg--Wilson theory.
Reference~\onlinecite{jiang19} found close parallels between this transition and an easy-plane Neel to VBS DQCP in 2d, but the 1d version is more tractable and, in particular, allows a controlled field theory description.
Such a transition therefore is proposed to constitute an example of deconfined quantum criticality in 1d \footnote{There are also 1D fermionic models with non-Landau continuous transitions, which were studied in Refs.~\onlinecite{sengupta2002, sandvik2004}.}.
While some of the properties of this DQCP are special to 1d, such as continuously varying critical indices, the field theory description offers new perspectives that could be worthwhile to pursue in 2d as well~\cite{jiang19}.

Here, we present strong numerical evidence supporting the 1d proposal in a concrete model.
We use matrix product states (MPS) working directly in the thermodynamic limit, and we develop a specialized ``finite-entanglement scaling'' protocol that allows us to study this transition with high precision.
A nontrivial aspect of the infinite-volume MPS study of the DQCP is that the MPS ground state at fixed bond dimension undergoes a first-order transition, which turns out to be advantageous for accessing properties of the true continuous DQCP via scaling in finite bond dimension.
Our numerical study confirms key predictions of the 1d DQCP theory, thus providing a definitive example of such a phase transition.

The paper is organized as follows.
In Sec.~\ref{sec:description}, we give an overview of the system and symmetries, summarize field theory predictions for the transition, and introduce our concrete model and its phase diagram.
In Sec.~\ref{sec:mps_study}, we describe the numerical study of the ferromagnet to VBS transition, including details of our finite-entanglement scaling protocol which leads to an accurate determination of the critical indices, and study the variation along the phase boundary.
In Sec.~\ref{sec:coex}, we study the regime where the transition splits into two, with an intervening phase of coexistence of magnetic and VBS orders.
We conclude in Sec.~\ref{sec:conclusion} with discussion of possible future directions.
We also include three appendices: Appendix \ref{app:meanfield_sep} provides a basic mean-field description of the phase diagram using pictures of the ground states described by separable wavefunctions.
Appendix \ref{app:meanfield_mps} resolves some questions arising from the separable-state mean-field picture by representation of model ground states for the phases as analytic MPS of bond dimension two.
Finally, Appendix \ref{app:delta1} develops a field theory description of another phase transition encountered in this model beyond the LGW symmetry-breaking paradigm.

\section{Description of model}
\label{sec:description}

Here we summarize the key results of Ref.~\onlinecite{jiang19}, which contains a number of descriptions of the model at hand.
Briefly, a second-order phase transition was proposed at the phase boundary of an Ising ferromagnet and valence bond solid (VBS).
Because these states break different symmetries, a continuous phase transition between them falls outside of the Landau--Ginzburg--Wilson paradigm.

\subsection{General model and symmetries}
\label{subsec:model}

Our general Hamiltonian is the following spin model, with nearest- and next-nearest-neighbor terms:
\begin{equation}
\begin{aligned}
H = \sum_j \Big( -\,& J_x \sigma^x_j \sigma^x_{j+1} - J_z \sigma^z_j \sigma^z_{j+1} \\
&+ K_{2x} \sigma^x_j \sigma^x_{j+2} + K_{2z} \sigma^z_j \sigma^z_{j+2} \Big) ~.
\end{aligned}
\label{eq:H}
\end{equation}
We take $J_x$, $J_z$, $K_{2x}$, $K_{2z}$ nonnegative, that is, with ferromagnetic nearest-neighbor and antiferromagnetic next-nearest-neighbor interactions.
$H$ respects two Ising-like symmetries as well as time reversal:
\begin{align}
g_x &= \prod_j \sigma^x_j~: \quad \sigma^x_j \mapsto \sigma^x_j~,~ \sigma^{y,z}_j \mapsto -\sigma^{y,z}_j~; \\
g_z &= \prod_j \sigma^z_j~: \quad \sigma^z_j \mapsto \sigma^z_j~,~ \sigma^{x,y}_j \mapsto -\sigma^{x,y}_j~; \\
\mathcal T &= \left(\prod_j i \sigma^y_j \right) \mathcal K~: \quad \sigma^\alpha_j \mapsto -\sigma^\alpha_j~,~ i \mapsto -i ~.
\end{align}
Here $\mathcal K$ is complex conjugation in the $\sigma^z$ basis.
The model also has translation symmetry, $T_1: \sigma^\alpha_j \mapsto \sigma^\alpha_{j+1}$, as well as inversion symmetry $I: \sigma^\alpha_j \mapsto \sigma^\alpha_{-j+1}$, which we take to be about a bond center.

In the regime where $J_z$ is dominant, the spins order as a ferromagnet in the $\sigma^z$ direction; we call this phase ``\zFM.''
For intermediate $K_{2x} \sim K_{2z}$, the spins are disordered (all on-site symmetries are preserved) and instead form dimers on alternating bonds; we call this phase ``\VBSI,'' to distinguish from other specific dimer states which we encounter.
A fixed-point picture of this particular VBS phase is a product state of dimers on, say, all $(2m-1, 2m)$ bonds, where each dimer is an entangled state of two spins of the form
\begin{equation}
\begin{aligned}
\ket{D^{(I)}_{12}} & = \frac{ \ket{+\hat z}_1 \ket{+\hat z}_2 + \ket{-\hat z}_1 \ket{-\hat z}_2 }{\sqrt 2} \\
& = \frac{ \ket{+\hat x}_1 \ket{+\hat x}_2 + \ket{-\hat x}_1 \ket{-\hat x}_2 }{\sqrt 2} \\
& = \frac{ \ket{+\hat y}_1 \ket{-\hat y}_2 + \ket{-\hat y}_1 \ket{+\hat y}_2 }{\sqrt 2} ~.
\end{aligned}
\label{eq:d12-I}
\end{equation}
Note that the spins in the dimer have ferromagnetic $zz$ and $xx$ correlations and antiferromagnetic $yy$ correlations.
This state is expected from the ferromagnetic $J_z$ and $J_x$ couplings.
Most of the time, we will focus on the \VBSI phase and will frequently refer to it as simply \VBS where it does not cause confusion.

The above fixed-point VBS wavefunction is an exact ground state at the Majumdar--Ghosh point: $J_x = J_z = J$, $K_{2x} = K_{2z} = K_2$, and $K_2/J = 0.5$~\cite{MajumdarGhosh1969, MajumdarGhosh1969II, FurukawaSatoFurusaki2010, FurukawaSatoOnodaFurusaki2012}.
Our primary focus is on the phase transition between the \zFM and \VBSI phases.

\subsection{Summary of field theory for the \texorpdfstring{\zFM}{zFM} to \texorpdfstring{\VBS}{VBS} transition}
\label{subsec:ft}

The field theory description of the \zFM to \VBS transition in Ref.~\onlinecite{jiang19} has a Luttinger liquid--like form and is written in terms of conjugate fields $\phigood$ and $\thtgood$, with velocity $\tilde{v}$ and Luttinger parameter $\tilde{g}$:
\begin{align}
S[\phigood, \thtgood] \!=\! & \int\! \dd \tau\, \dd x \left[ \frac{\ii}{\pi} \partial_\tau \phigood \partial_x \thtgood + \frac{\tilde{v}}{2\pi} \! \left( \frac{1}{\tilde{g}}(\partial_x \thtgood)^2 + \tilde{g} (\partial_x \phigood)^2 \right) \right] \nonumber \\
&+ \int\! \dd \tau\, \dd x \left[ \lambda \cos(2\thtgood) + \lambda' \cos(4\thtgood) + \kappa \cos(4\phigood) \right].
\label{eq:gauged_bosonization_action}
\end{align}
The notation here matches that in Ref.~\onlinecite{jiang19} (see Sec.~VII there); in particular, tildes over the fields signify that they are not simply related to a naive bosonization of spins in the $xz$ plane.

As written, the fields have periodicities $\phigood + \pi \equiv \phigood$ and $\thtgood + 2\pi \equiv \thtgood$, which follows from their partonic origin (see Sec.~VII in Ref.~\onlinecite{jiang19} for details and also App.~E there for another perspective on this theory).
The second line shows the leading symmetry-allowed cosine terms of the fields.
Taking the Luttinger parameter in the range $\tilde{g} \in (1/2, 2)$ arranges that the $\lambda'$ and $\kappa$ terms are irrelevant and the $\lambda$ term is the only relevant cosine.
The \zFM to VBS transition occurs when the relevant coupling $\lambda$ changes sign, hence the critical theory is Gaussian.
The correlation length exponent follows from the scaling dimension of the relevant cosine perturbation and is given by
\begin{align}
\nu = \frac{1}{2 - \tilde{g}} ~,
\label{eq:nu}
\end{align}
which can vary in the range $\nu \in (2/3, \infty)$ for $\tilde{g} \in (1/2, 2)$.

The most important observables are the \zFM and VBS order parameters, which are given by
\begin{align}
M_z^\text{FM} \sim \sin(\thtgood) ~,~ \quad
\Psi_{\VBS} \sim \cos(\thtgood) ~.
\end{align}
At the critical point, they have the same scaling dimension
\begin{align}
\text{dim}[M_z^\text{FM}] = \text{dim}[\Psi_{\VBS}] = \frac{\tilde{g}}{4} ~,
\label{eq:scalingdims_zFM_VBS}
\end{align}
which can vary in the range $(1/8, 1/2)$.
The scaling dimension of an observable $\O$ determines the power law decay of the critical correlations: if $\lrangle{\O(x) \O(0)} \sim 1/x^{p_\O}$, then $p_\O = 2\, \text{dim}[\O]$.
General scaling arguments also yield the order parameter onset exponent $\beta = \nu\, p/2$.

We also mention the next most prominent observables, namely the \xFM and \yAFM order parameters
\begin{align}
M_x^\text{FM} \sim \cos(2\phigood) ~,~ \quad
M_y^\text{AFM} \sim \sin(2\phigood) ~,
\end{align}
with scaling dimensions
\begin{align}
\text{dim}[M_x^\text{FM}] = \text{dim}[M_y^\text{AFM}] = \frac{1}{\tilde{g}}~,
\label{eq:dim_xFM_yAFM}
\end{align}
which can vary between $2$ and $1/2$.
Note that the dominant $\sigma^x$ correlations are ferromagnetic while the dominant $\sigma^y$ correlations are antiferromagnetic.
This is tied to the fact that this theory describes the transition from the \zFM phase to the \VBSI phase with the fixed-point elementary dimer given by Eq.~\eqref{eq:d12-I}; see also the discussion following that equation.

To summarize, the critical exponents vary continuously and depend on a single parameter $\tilde{g}$.
When $\tilde{g}$ drops below $1/2$, the $\lambda'$ term becomes relevant and destabilizes the above picture for the direct transition between the \zFM and VBS phases.
Analysis in Ref.~\onlinecite{jiang19} suggests that for $\lambda' > 0$, an intermediate phase with coexisting \zFM and VBS order parameters appears between the pure \zFM and pure VBS phases.
We will also examine this scenario in our study of the specific model below.

\subsection{Specific model and expected phase behavior}
\label{subsec:pt}

In order to study the phase transition between the Ising ferromagnet and VBS phases, we restrict in parameter space to a two-dimensional slice given by $K_2 = K_{2x} = K_{2z}$ and $\delta = (J^z - J^x)/(J^z + J^x)$; that is, $J^z = J(1 + \delta)$ and $J^x = J(1 - \delta)$, where we will take $J = 1$.
The $U(1)$ symmetry of rotations in the $xz$ plane is broken only by the nearest-neighbor couplings, and is restored for anisotropy $\delta = 0$.
The point $\delta = K_2 = 0$ is the XX model, which maps to free fermions and belongs to the quasi--long-range-ordered (QLRO) phase present on the $\delta = 0$ axis up to some critical $K_{2, \text{KT}}$.
Along this axis at $K_{2, \text{KT}}$, the model undergoes a Kosterlitz--Thouless transition~\cite{KosterlitzThouless1973, Kosterlitz1974, Haldane1982} to the VBS phase described earlier.
Additional phases occur at significantly larger $K_2$ and were studied in Refs.~\onlinecite{FurukawaSatoFurusaki2010, FurukawaSatoOnodaFurusaki2012} but are not considered in the present work.
For any $|\delta| > 0$, at small values of $K_2$ the ground state is an Ising ferromagnetic state.
At intermediate $K_2$ the VBS phase is stable to introducing spin anisotropy and extends to non-zero $\delta$.
At fixed finite $\delta$, we therefore expect that increasing the $K_2$ term from small values will drive a transition from the Ising ferromagnet to the VBS phase.

It is sufficient to consider $\delta \geq 0$, as the Hamiltonian with parameters $\{-\delta, K_2\}$ is equivalent to that with $\{\delta, K_2\}$ up to a local unitary rotation, $\sigma_j^x \mapsto \sigma_j^z, \sigma_j^z \mapsto -\sigma_j^x$, which takes the \zFM phase to an Ising $x$ ferromagnet (``\xFM''), and vice versa.
This transformation leaves the \VBSI dimer of Eq.~\eqref{eq:d12-I} invariant, thus the same phase appears for both positive and negative $\delta$.

We may also restrict our focus to models with $\delta \leq 1$ due to another relationship---namely, that models having parameters $\{\delta, K_2\}$ and $\{\delta', K_2'\} = \{1/\delta, K_2/\delta\}$ are related by local unitary $\prod_m \sigma^z_{2m}$, taking $\sigma^z_{2m} \mapsto \sigma^z_{2m}$ and $\sigma^x_{2m} \mapsto -\sigma^x_{2m}$.
Indeed, the primed model has parameter values $J^{z\prime} = 1 + \delta' = (1 + \delta)/\delta$, $J^{x\prime} = 1 - \delta' = -(1 - \delta)/\delta$, and the given rotation relates it to the first model up to an overall energy scale.
Note that under this unitary transformation, the elementary dimer wavefunction Eq.~\eqref{eq:d12-I} maps to
\begin{equation}
\begin{aligned}
\ket{D_{12}^{(II)}} & = \frac{ \ket{+\hat z}_1 \ket{+\hat z}_2 - \ket{-\hat z}_1 \ket{-\hat z}_2 }{\sqrt 2} \\
& = \frac{ \ket{+\hat x}_1 \ket{-\hat x}_2 + \ket{-\hat x}_1 \ket{+\hat x}_2 }{\sqrt 2} \\
& = \frac{ \ket{+\hat y}_1 \ket{+\hat y}_2 + \ket{-\hat y}_1 \ket{-\hat y}_2 }{\sqrt 2} ~.
\end{aligned}
\label{eq:d12-II}
\end{equation}
Hence, at values $\delta > 1$ one finds another dimer state, which we denote ``\VBSII,'' as it is a distinct phase from the previously described \VBSI provided the on-site symmetries are not broken~\cite{jiang19}.
The precise distinction between the phases is that on a periodic system with an odd number of dimers, the ground states in \VBSI have quantum numbers $(g_x, g_y, g_z) = (1, -1, 1)$, whereas the quantum numbers in \VBSII are $(g_x, g_y, g_z) = (-1, 1, 1)$.

Naively, one may expect a phase transition between \VBSI and \VBSII at $\delta = 1$.
As we discuss in Sec.~\ref{sec:delta1}, the actual situation in this model is somewhat more complicated: in a particular region of the phase diagram close to the \zFM phase, the spin system also develops \zFM order on top of \VBSI or \VBSII, and this coexisting broken on-site symmetry allows a continuous connection between the two dimer states.
Finally, for larger $K_2$, another phase---which does not appear in the field theory---arises in our model intervening between the two dimer phases.
This is the so-called ``up-up-down-down'' state in the $\sigma^x$ basis, or ``$x$UUDD.''
The ground state of this phase breaks $T_1$ and $g_z$ and has the following fixed-point wavefunction:
\begin{equation}
\ket{x\text{UUDD}} = \otimes_n \ket{+\hat x}_{4n-3}\, \ket{+\hat x}_{4n-2} \, \ket{-\hat x}_{4n-1} \, \ket{-\hat x}_{4n}~.
\end{equation}

In App.~\ref{app:meanfield_sep}, we give fixed-point pictures and mean field energetics for all phases encountered in our window of study, thus providing some intuition for the observed phase diagram.

\section{Study of \texorpdfstring{\zFM}{zFM} to \texorpdfstring{\VBS}{VBS} phase transition}
\label{sec:mps_study}

We make use of the recently-developed numerical method ``variational uniform matrix product states'' (VUMPS), which is similar to infinite-system DMRG (IDMRG) but has been demonstrated to achieve superior convergence in some cases \cite{zauner18}.
Like IDMRG, this method optimizes over MPS in the thermodynamic limit; that is, the ansatz is specified by a finite set of tensors comprising the unit cell of the wavefunction, which contain the variational parameters of the infinite state.
The understanding of VUMPS is geometrical: one searches within the manifold of uniform MPS of fixed bond dimension for the point $\ket{\psi^\ast}$ at which the energy residual $(H-E)\ket{\psi^\ast}$ is orthogonal to the manifold.
This optimization can be formulated in the ``post-MPS'' tangent space language \cite{haegeman13}, but turns out to be similar to IDMRG.

\begin{figure}[ht]
\centering
\includegraphics[width=\columnwidth]{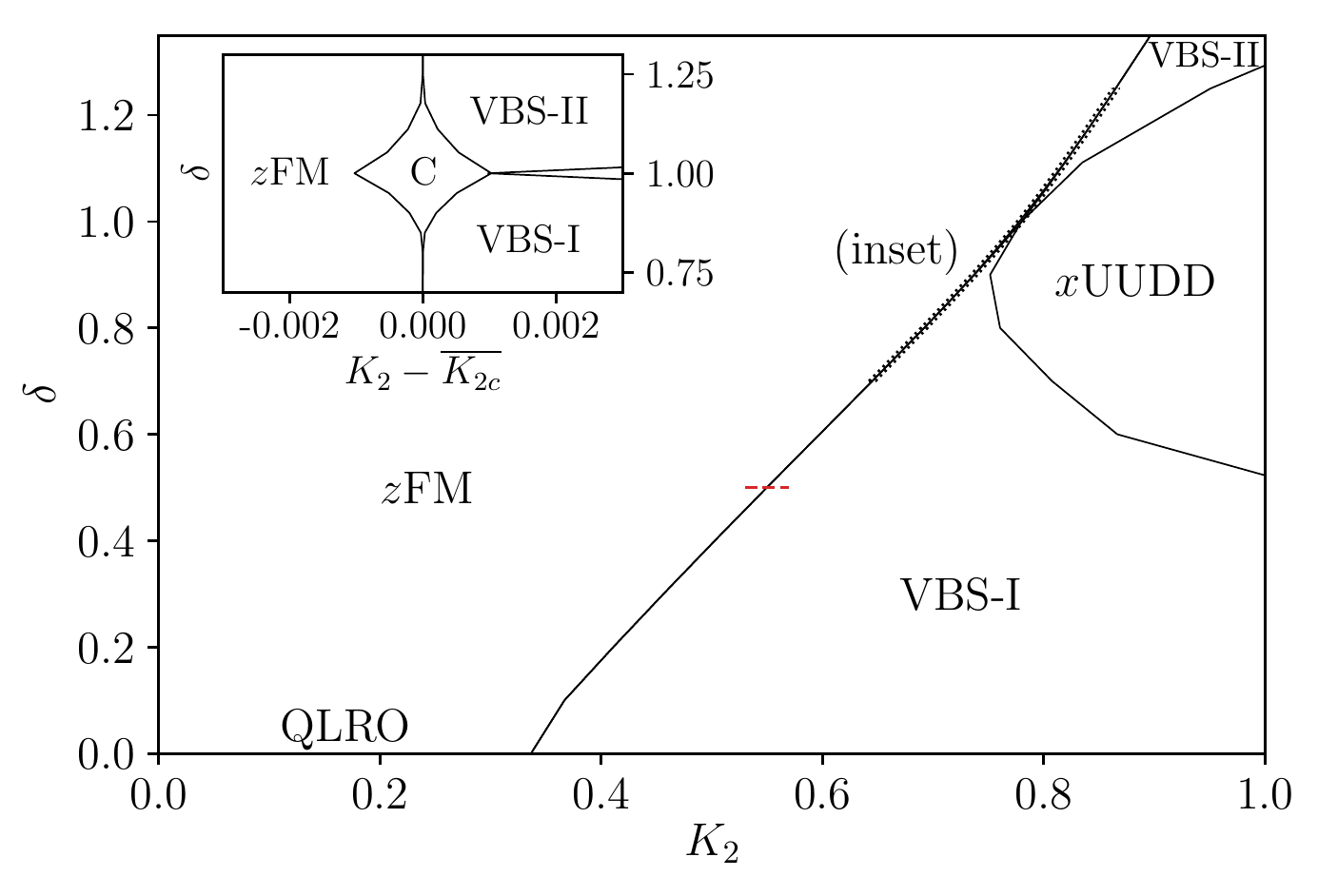}
\caption{\label{fig:phase_diagram}
The phase diagram in the $K_2$-$\delta$ plane includes the \zFM, \VBSI, \VBSII, and $x$UUDD phases.
Inset shows a centered view of the coexistence region, denoted ``C'', appearing between the \zFM and \VBSI or \VBSII phases for $\delta$ close to $1$.
While the distinction between the \VBSI and \VBSII phases is protected by the on-site symmetries, the \VBSI+\zFM and \VBSII+\zFM coexistence phases are not distinct and there is no transition inside the C region.
The cut indicated at $\delta=0.5$ will be investigated in detail in Sec.~\ref{subsec:critical_point} as an example case.
}
\end{figure}

The uniform MPS ansatz actually provides a dressed mean-field description of the phase transition~\cite{liu10}.
Because the mean-field treatment in the present case exhibits a first-order phase transition, one expects the VUMPS method to encounter metastability effects near the phase transition arising from competing orders.
We describe our protocol to address this challenge below; we aer in fact able to utilize the first-order behavior of the finite--bond dimension MPS to make very accurate determinations of the phase boundary.
We first show in Fig.~\ref{fig:phase_diagram} our result for the phase diagram outlined in Sec.~\ref{subsec:pt}, and in the following sections we provide a methodological description.

\subsection{Representative study along \texorpdfstring{$\delta = 0.5$}{delta = 0.5} cut}
\label{subsec:critical_point}

We illustrate our method of studying this phase transition by  discussing in detail a concrete cut through the phase diagram, namely along the line $\delta = 0.5$ generated by varying the parameter $K_2$.
Afterward, we will generalize to obtain a full description of the phase boundary by repeating the same process for multiple slices at constant $\delta$.
The line at $\delta = 0.5$ is generic, having no symmetries additional to those specified in Sec.~\ref{sec:description}.
This slice is indicated in Fig.~\ref{fig:phase_diagram}.

\subsubsection{Broad description of phase transition}
\label{subsubsec:broad_pt}

\begin{figure}[ht]
\centering
\includegraphics[width=\columnwidth]{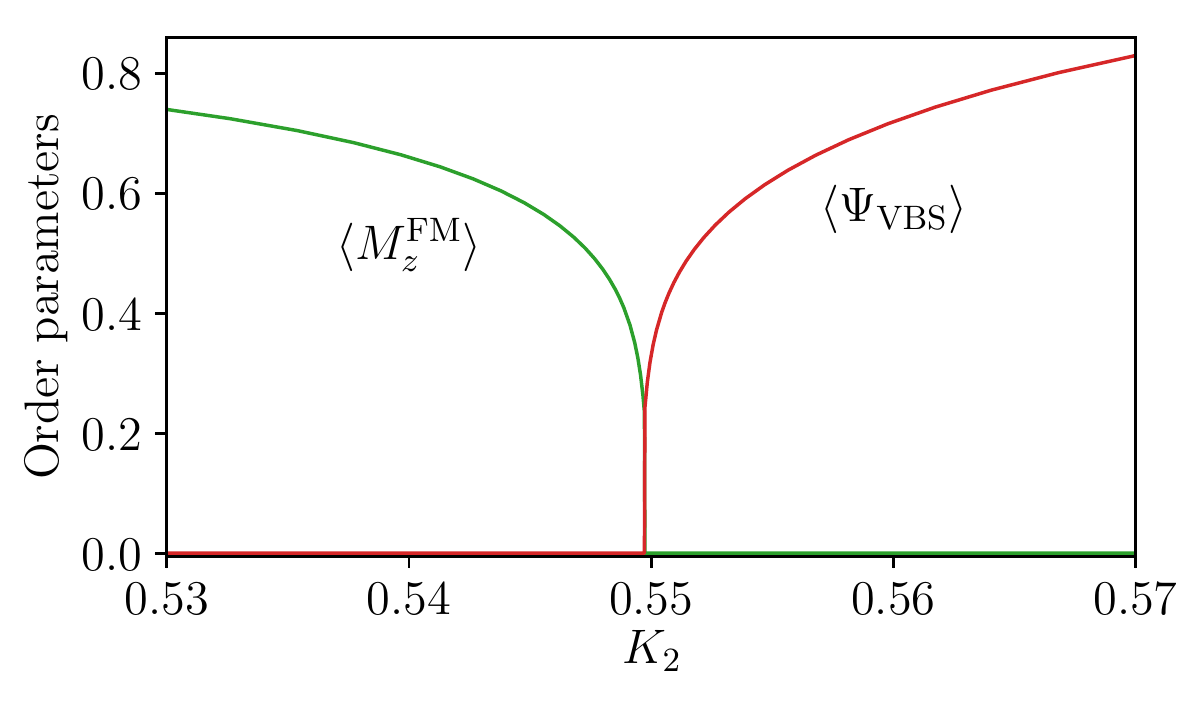}
\caption{\label{fig:op_d0.5} (color online)
Transition between the \zFM and VBS phases, as detected by the corresponding order parameters.
This scan is taken at fixed $\delta = 0.5$ using bond dimension $\chi = 192$.
We observe that at fixed $\chi$, the MPS ground state shows a first-order transition; the discontinuities in the order parameters decrease towards zero with increasing $\chi$, as studied in detail in Fig.~\ref{fig:beta_d0.5}.}
\end{figure}

One can attain a basic understanding of the phase transition via simple analysis of the optimized MPS ground states.
Using ansatz trial states originating within each phase, we tune $K_2$ through the critical point and observe the evolution of certain properties of the trial state wavefunction.
The most evident indication of the phase transition is the order parameter for each phase acquiring a finite expectation value.
Because the numerical method preferentially selects states of low entanglement, it finds everywhere a representative of the ground state manifold with spontaneously broken symmetry.
As both phases break $\Z_2$ symmetries ($g_x$ in the \zFM phase and $T_1$ in the VBS phase), both ground state degeneracies are two and the symmetry breaking manifests as a sign in the expectation value of the corresponding order parameters.
The order parameter for the \zFM phase is
\begin{equation}
\lrangle{M_z^\text{FM}} = \lrangle{\sigma^z_0} ~,
\end{equation}
where the site label $0$ indicates the first tensor in the unit cell, which in this case has only a single site.
For the VBS phase, the order parameter is
\begin{equation}
\lrangle{\Psi_\text{VBS}} = \lrangle{{\bm \sigma}_0 \cdot {\bm \sigma}_1 - {\bm \sigma}_1 \cdot {\bm \sigma}_2} ~,
\end{equation}
where ${\bm \sigma}_j$ denotes the Pauli vector acting at site $j$.
The ground state of this phase has a two-site unit cell.
We ignore the sign in both order parameters, always implicitly taking the absolute value.

\begin{figure}[ht]
\centering
\includegraphics[width=\columnwidth]{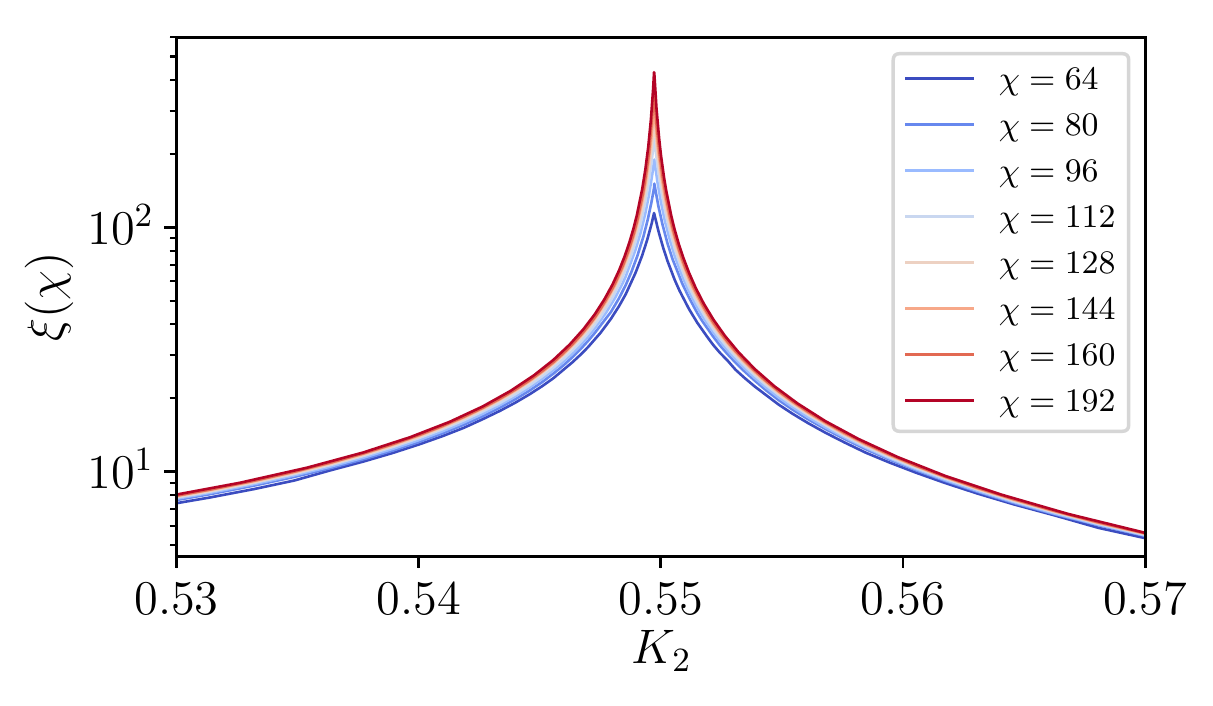}
\caption{\label{fig:xi_d0.5} (color online) 
The divergence of the correlation length in the exact ground state at the critical point manifests as a $\chi$-dependent cusp in the MPS correlation length $\xi(\chi)$; specifically, the height grows as a power law with $\chi$, as studied in detail in Fig.~\ref{fig:nu_d0.5}.
This feature is indicative of a continuous phase transition.
}
\end{figure}

The order parameters are shown in Fig.~\ref{fig:op_d0.5} for a large bond dimension $\chi = 192$.
As suggested by the mean field analysis, we do in fact find a discontinuous transition, with sizable jumps in both order parameters.
However, we argue that the true transition in the $\chi \to \infty$ limit is continuous.
Moreover, we use the first-order nature of the finite-$\chi$ approximants to our advantage: in particular, we will understand how the size of the order parameter discontinuity scales to zero with increasing $\chi$.

\begin{figure}[ht]
\centering
\includegraphics[width=\columnwidth]{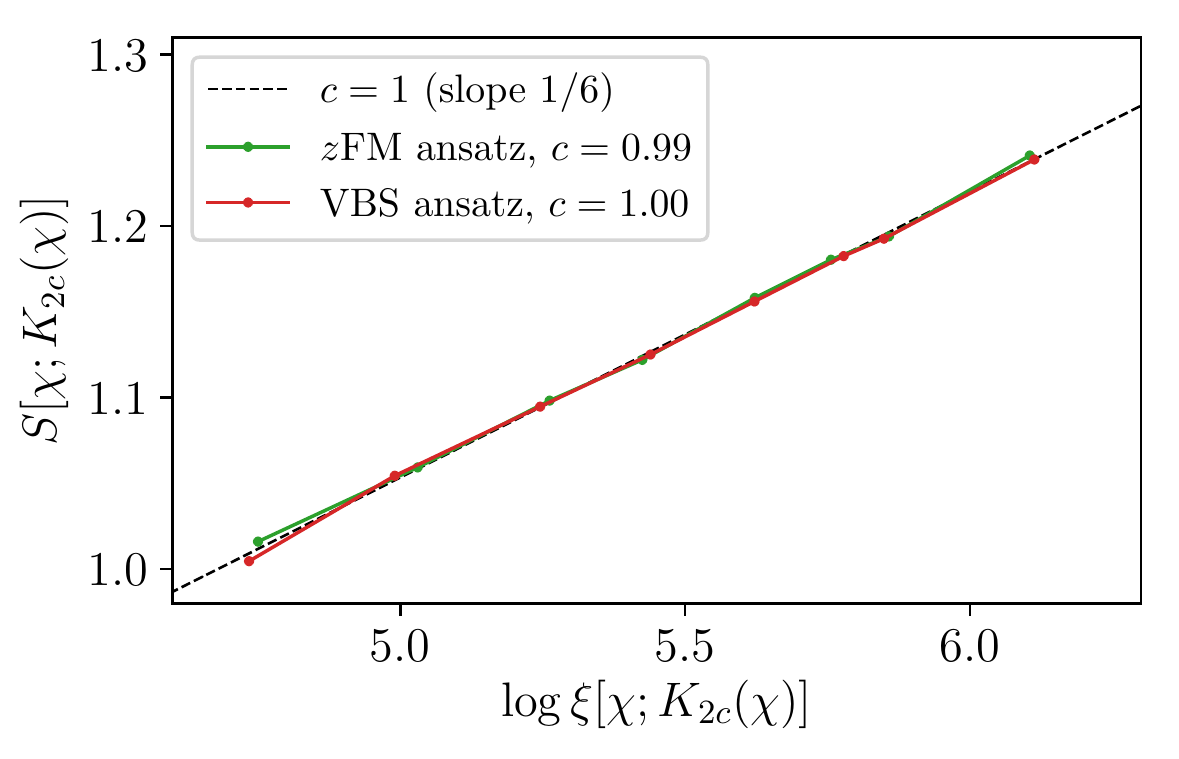}
\caption{\label{fig:c_d0.5} (color online) The scaling of the critical entanglement entropy $S[\chi;K_{2c}(\chi)]$ is nearly linear in $\log \xi[\chi;K_{2c}(\chi)]$, with the slope in good agreement with predicted central charge $c=1$.
Data shown is taken at parameter $\delta = 0.5$, and the dashed line is provided as a guide to the eye.
The pseudocritical point $K_{2c}(\chi)$ is defined later in the text and included here only for specificity; it is important insofar as it is particular to the MPS of bond dimension $\chi$.
}
\end{figure}

Another fundamental characterization of the phase transition is the behavior of the correlation length $\xi(\chi)$ of the minimum-energy state on the manifold of MPS of bond dimension $\chi$.
This quantity is a property of the spectrum of the MPS transfer matrix $T$.
In the simplest case of a single-site unit cell, $T = \sum_\sigma A^{\dag\sigma} \otimes A^\sigma$, where $\sigma$ runs over a basis of the local Hilbert space.
Normalization constrains the largest eigenvalue to be unity; the MPS correlation length is set by the second-largest eigenvalue, which dictates the slowest decay possible in the state.
Specifically, if $T$ spans a unit cell of $n$ sites, then $\lambda_2/\lambda_1 = e^{-n/\xi(\chi)}$, or $\xi(\chi) = -n/\log \lambda_2$.

\begin{figure}[bt!]
\centering
\includegraphics[width=\columnwidth]{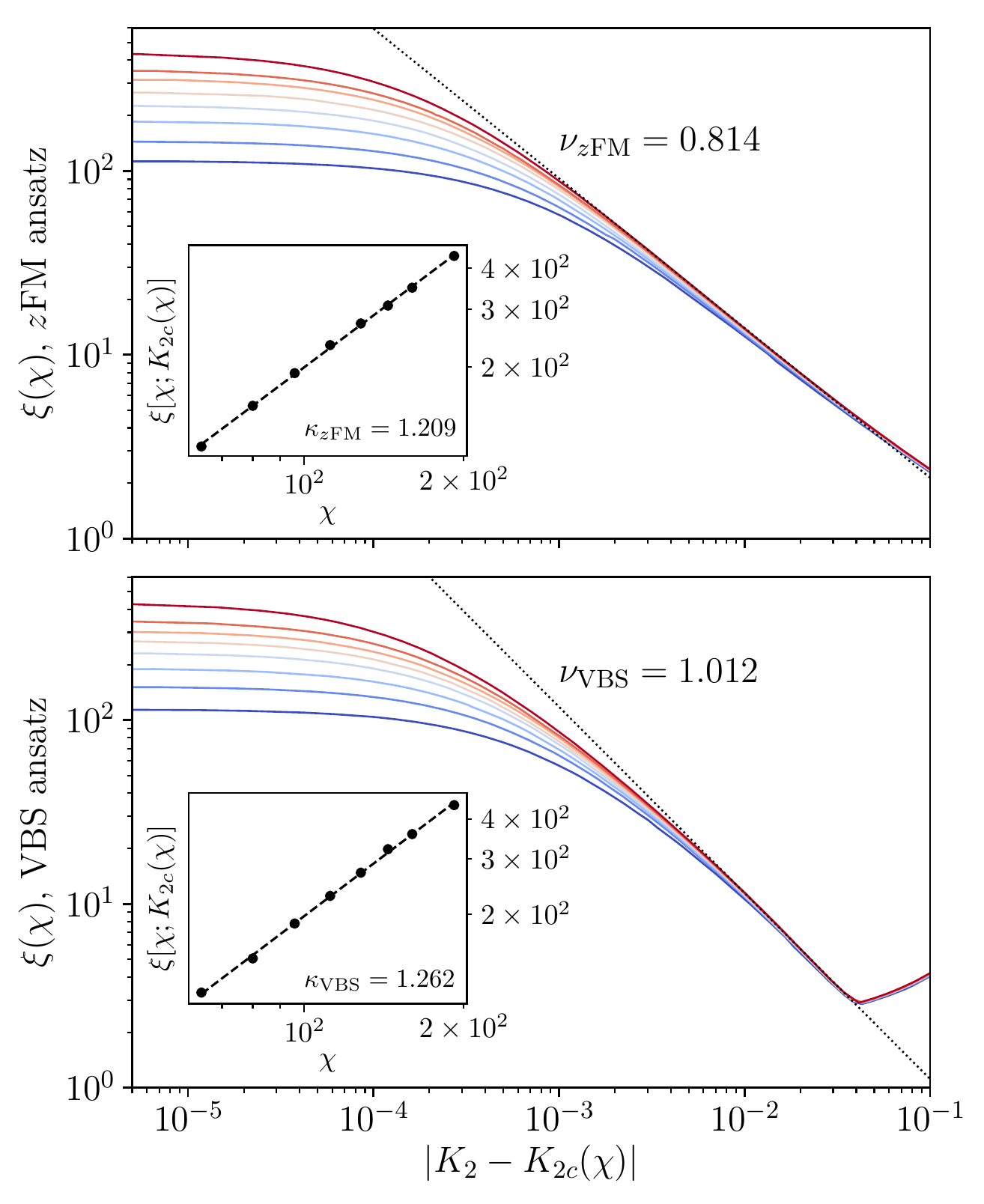}
\caption{\label{fig:nu_d0.5} (color online)
The MPS correlation length $\xi(\chi)$ exhibits power-law behavior in an intermediate region around $K_{2c}(\chi)$, here shown in the \zFM phase in the top panel and \VBS in the bottom.
Close to the pseudocritical point, the correlation length saturates to a maximum value dependent on the bond dimension, whereas farther away it approaches a constant in the gapped phase.
In the case of the \VBS phase, a nearby critical point (the transition to the $x$UUDD phase) affects the behavior of $\xi(\chi)$.
In the insets, we show the dependence of the maximum correlation length $\xi[\chi; K_{2c}(\chi)]$ extrapolated to the pseudocritical point $K_{2c}(\chi)$ as a function of $\chi$.
A fit to the scaling form Eq.~\eqref{eq:xi_vs_chi} is shown (note that the axes are logarithmic), along with extracted values of $\kappa$.
}
\end{figure}

We use $\xi$ without an argument to refer to the correlation length of the ground state and use $\xi(\chi)$ for the MPS correlation length.
At a continuous phase transition, the true correlation length $\xi$ diverges; however $\xi(\chi)$ remains finite, as $\lambda_2 < \lambda_1$ by injectivity.
Nevertheless, inside a gapped phase $\xi(\chi) \to \xi$, and where $\xi$ diverges $\xi(\chi)$ exhibits a cusp with $\chi$-dependent height.
We discuss this relationship further in Sec.~\ref{subsubsec:precise_pt}.
The MPS correlation length at the \zFM to VBS phase transition is shown in Fig.~\ref{fig:xi_d0.5}, and indeed displays a strong $\chi$-dependent cusp at the critical point.
At our largest $\chi = 192$, $\xi(\chi)$ already exceeds 400 lattice spacings, with consistent growth in $\chi$ (see our later study in Fig.~\ref{fig:nu_d0.5}).
This is the first strong evidence of a second-order transition, despite the order parameter discontinuity observed at this $\chi$.

As further evidence for a second-order transition, Fig.~\ref{fig:c_d0.5} shows the entanglement entropy in the optimized MPS versus the logarithm of the MPS correlation length near criticality.
For each $\chi$ we show two data points, measured in both the ansatz originating in the \zFM  and \VBS phases, each tuned to a point still in the phase but very close to the MPS transition at this $\chi$.
The relationship is consistent with the finite-entanglement scaling form~\cite{pollmann09}
\begin{equation}
S(\chi) = \frac{c}{6} \log \xi(\chi) ~,
\end{equation}
where $c$ is the central charge of the critical system.
The central charge estimates from fits to the above form are given in the figure and are consistent with the expected $c = 1$ from the theory of the \zFM to \VBS transition.

\subsubsection{Precise identification of critical point}
\label{subsubsec:precise_pt}

\begin{figure}[ht]
\centering
\includegraphics[width=\columnwidth]{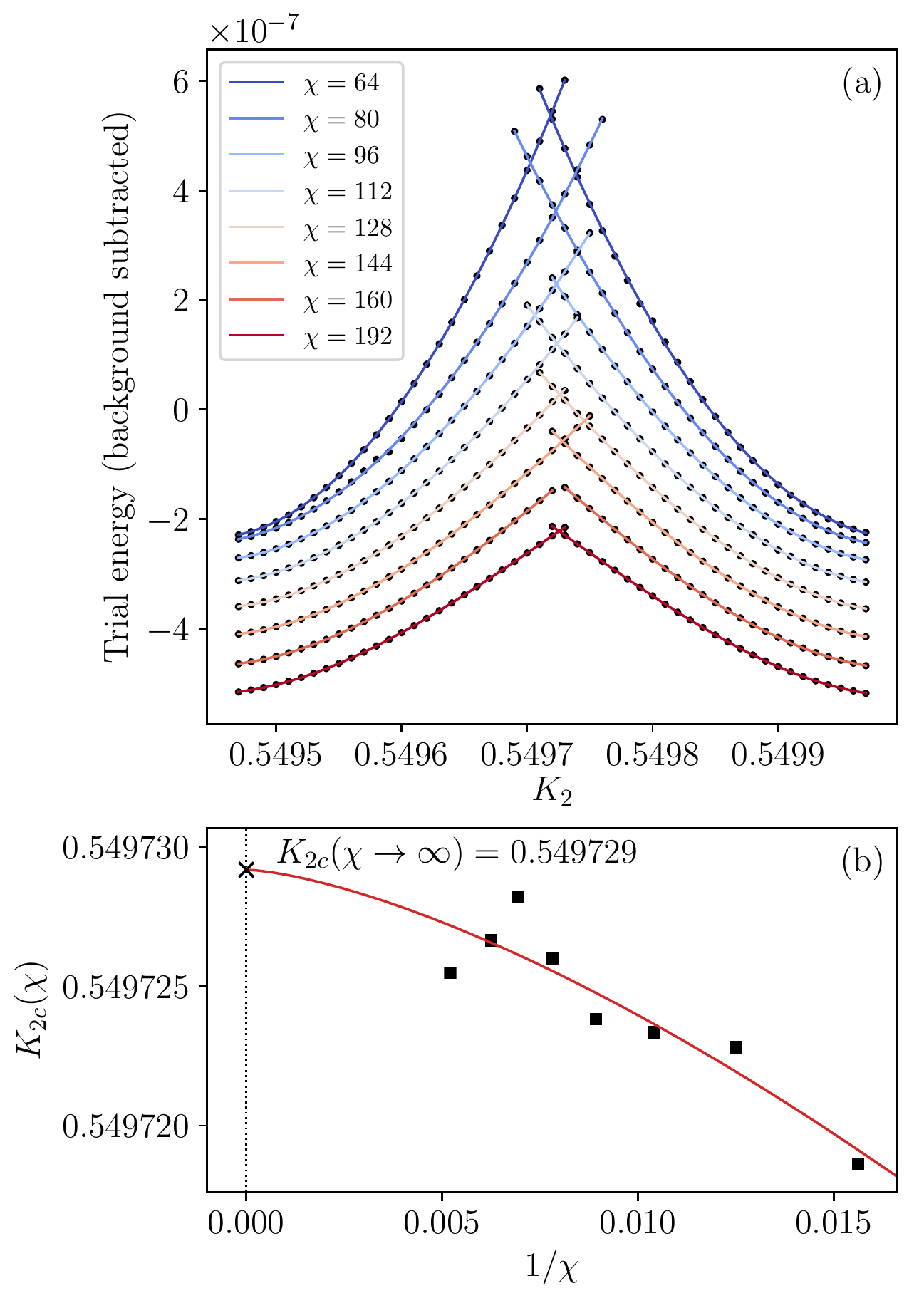}
\caption{\label{fig:K2c_d0.5} (color online)
Illustration of the process of locating the critical point from finite-entanglement scaling at $\delta = 0.5$.
(a) The energies of both trial wavefunctions from the \zFM and \VBS phases (fully optimized at each $K_2$) follow smooth curves, which determine the level crossing for a given bond dimension $\chi$ to a finer resolution than the scan in parameter $K_2$ via interpolation.
Due to hysteresis, in many cases we directly observe the crossing using the adiabatic protocol described in the text.
(b) Using the finite-entanglement scaling form Eq.~\eqref{eq:K2c_scal}, we extrapolate from the extracted pseudocritical $K_{2c}(\chi)$ to estimate the location of the critical point at $\chi \to \infty$.
The scatter in data points is not noise from the variational algorithm, but rather may be a consequence of the uneven spacing of the entanglement spectrum.}
\end{figure}

In principle, MPS methods are not well suited for describing ground states of quantum systems tuned to critical points, as the high degree of entanglement places a strong constraint on the accuracy of MPS (``classical'') approximations.
In contrast, ground states of gapped phases are well represented by MPS; however, in practice one can hope only to approach sufficiently close to a continuous phase transition to observe its true critical behavior.
Beyond some crossover point set by the bond dimension, the MPS ground state instead flows to the phase transition described by the mean field theory of the model~\cite{liu10}.

While MPS are unable to directly access critical states, it turns out that in the present case we can take advantage of the fact that the mean field phase transition is discontinuous, as described in App.~\ref{app:meanfield_sep}, to accurately estimate the location of the critical point.
Until the crossover point the system exhibits the behavior of the true continuous phase transition, but in tuning the system through the critical point one instead observes a level crossing of states connected to the fixed-point descriptions of each phase.
In this regime the near-degeneracy of these dissimilar states leads to increased influence of the initial trial wavefunction in the VUMPS method, making convergence to the true ground state difficult when employing random initial states.
To circumvent this, we use an ``adiabatic'' protocol, first obtaining the MPS ground state in each phase far from the transition and slowly tuning the system to criticality in a series of discrete jumps, at each step allowing the state to converge fully.
Due to metastability effects, hysteresis develops very close to the critical point; however, we are always able to identify the true ground state from comparison of the trial state energies.
Because for MPS all energy levels are analytic functions of the Hamiltonian parameters, performing this scan in both phases allows one to identify the level crossing with a high degree of accuracy, in fact with a greater resolution than is used to tune the Hamiltonian.

This process is illustrated in Fig.~\ref{fig:K2c_d0.5} for a range of $\chi$, where in panel (a) we show the trial energies tracked from each side and in panel (b) we show the extracted locations of the level crossings as a function of $1/\chi$.
Note that the range of $K_2$ values is already very narrow, and the accuracy in the extrapolated crossings is better than $10^{-6}$.
Note also that the differences in the trial energies are enhanced by subtracting some smooth polynomial background (chosen for each $\chi$), and that the vertical scale is very small; the slope discontinuity in the VUMPS trial energy decreases towards zero with increasing $\chi$.

The above protocol applies to a uniform MPS having a fixed bond dimension $\chi$.
In fact, for any such ansatz with finite entanglement, the observed phase transition will occur not at the true critical point $K_{2c, \text{true}}$ but at some {\it pseudocritical} point $K_{2c}(\chi)$.
We expect that in the limit $\chi \to \infty$ the pseudocritical points converge to the true value.
\textcite{pollmann09} determined that for a critical system with infinite correlation length $\xi$, the correlation length of the minimum-energy MPS at fixed bond dimension scales as
\begin{equation}
\xi(\chi) \sim \chi^\kappa
\label{eq:xi_vs_chi}
\end{equation}
with exponent
\begin{equation}
\kappa = \frac{6}{c \left(\sqrt{\frac{12}{c}} + 1 \right)} ~,
\end{equation}
which depends on the central charge of the critical system.

In order to describe the dependence of the pseudocritical point on bond dimension, we adapt an argument from finite-size scaling in statistical mechanics, which is commonly used in Monte Carlo studies.
Denote the control parameter driving the transition as $h$, with the true critical point at $h_{c, \text{true}}$.
In a system of finite length $L$ the transition is smeared, but one can often identify a pseudocritical point $h_c(L)$ from some feature in the observables, such as peaks in susceptibilities, Binder ratio crossings, etc.
Finite-size scaling predicts that the pseudocritical points approach the true critical point as $h_c(L) - h_{c, \text{true}} \sim L^{-1/\nu}$, which follows from comparing the true correlation length at $h_c(L)$ with the length scale $L$ imposed by the system size.
We conjecture that similar relation holds for the infinite-system variational MPS study, by replacing $L$ with the length scale $\xi(\chi)$ imposed by the bond dimension:
\begin{equation}
K_{2c}(\chi) - K_{2c, \text{true}} \sim \xi(\chi)^{-1/\nu} \sim \chi^{-\kappa/\nu} ~.
\label{eq:K2c_scal}
\end{equation}
One can also imagine using this relation to extract the correlation length exponent $\nu$.
\footnote{In the 1d quantum Ising model studied in Ref.~\onlinecite{liu10}, the infinite-system MPS at fixed $\chi$ has a continuous mean field transition, and \eqref{eq:K2c_scal} provides a fairly accurate description of the approach of the corresponding pseudocritical points to the true critical point, with central charge $c = 1/2$ and correlation length exponent $\nu = 1$ for the Ising transition.}

Unfortunately, one observes in Fig.~\ref{fig:K2c_d0.5}(b) significant scatter in the values of $K_{2c}(\chi)$ on top of some smooth behavior.
This is not noise or evidence that the trial MPS is not energetically optimal, but rather a reproducible feature of the finite-$\chi$ results, which we conjecture arises from the nonuniformity of the gaps in the entanglement spectrum of the state.
The plotted curve and value of $K_{2c}(\chi \to \infty)$ was fitted by fixing the value of the correlation length exponent to $\nu \approx 0.914$ extracted from later analysis, and is presented primarily as a consistency check.
In any case, the $K_{2c}(\chi)$ vary over a very small range, and as our scaling analysis below involves only the pseudocritical points $K_{2c}(\chi)$, the uncertainty in $K_{2c}(\chi \to \infty)$ is irrelevant for our subsequent characterizations of the critical point.

\subsubsection{Correlation length and order parameter onset exponents}
\label{subsubsec:nu_beta_d0.5}

\begin{figure}[bt!]
\centering
\includegraphics[width=\columnwidth]{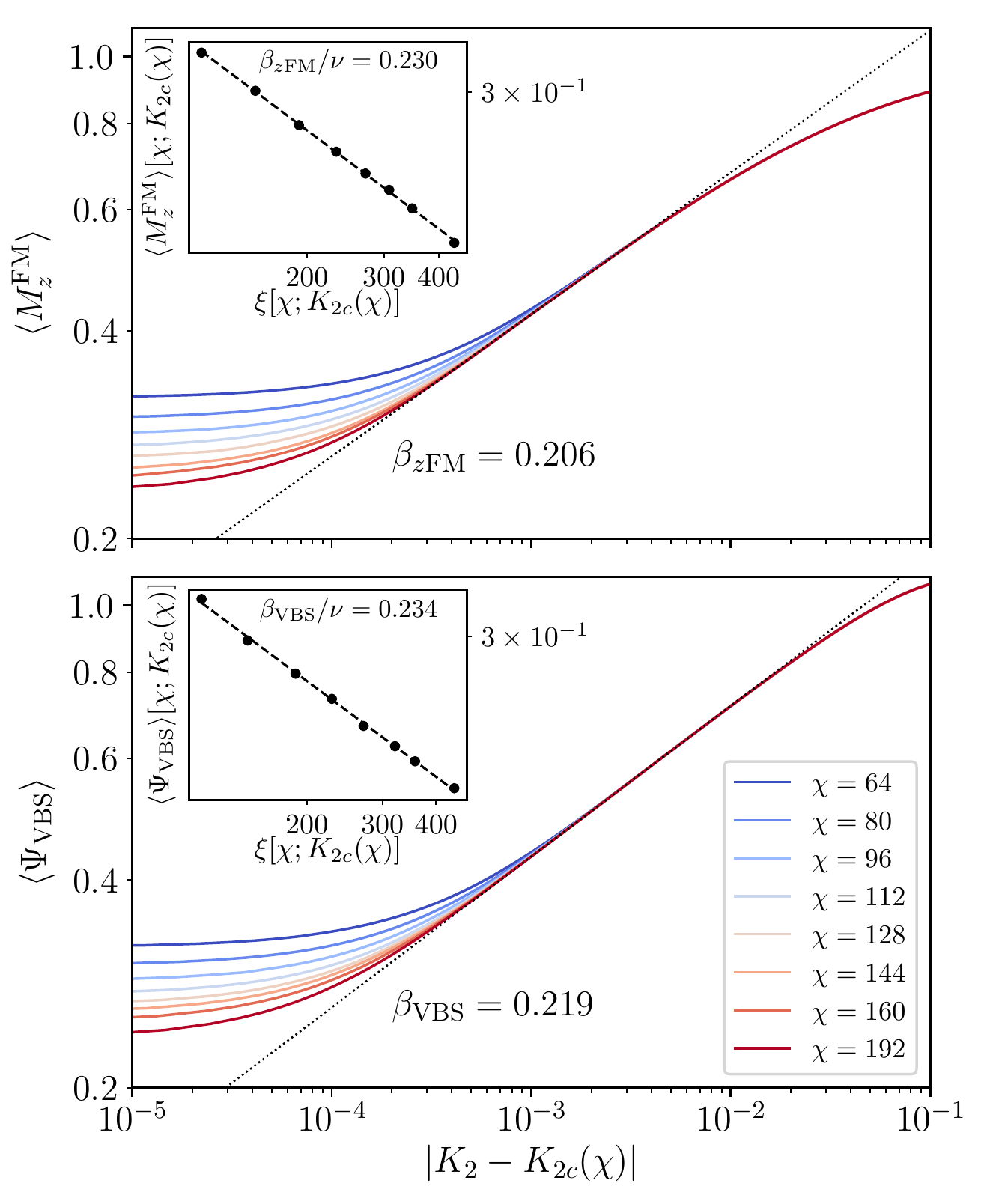}
\caption{\label{fig:beta_d0.5} (color online) 
Expectation values of the \zFM and VBS order parameters on the slice $\delta = 0.5$ show a region of power-law dependence in an intermediate range near the critical point which extends closer to the transition with increasing bond dimension.
Far from the critical point, the order parameters approach their maximal values, whereas very close to $K_{2c}(\chi)$ at fixed $\chi$, they saturate due to the discontinuous mean-field description of the transition.
The top panel shows $K_2$ in the \zFM phase, and the bottom panel $K_2$ in the VBS phase; in both panels, we give $K_2$ relative to the pseudocritical $K_{2c}(\chi)$ determined for each bond dimension as in Fig.~\ref{fig:K2c_d0.5}.
The dashed line in each panel shows the fitted power-law onset form with exponent $\beta_{\zFM}$ or $\beta_{\VBS}$ in this intermediate range, using the largest bond dimension data, which is essentially already converged to the infinite-$\chi$ values.
Insets show the  limiting values of the corresponding order parameters at $K_{2c}(\chi)$ as a function of the limiting $\xi[\chi, K_{2c}(\chi)]$, and a power law fit to Eq.~\eqref{eq:m_vs_xi}.
Note that inset axes use logarithmic scaling.
}
\end{figure}

Using the precise estimates of the finite-entanglement pseudocritical points from Sec.~\ref{subsubsec:precise_pt}, we are able to determine critical scaling exponents of the transition.
Specifically, we consider the correlation length exponent $\nu$ and the order parameter exponents for both phases $\beta_{\zFM}$ and $\beta_{\VBS}$.
The most straightforward way to determine $\nu$ is through its definition: $\xi \sim |K_2 - K_{2c}|^{-\nu}$.
In Fig.~\ref{fig:nu_d0.5} we show $\xi(\chi)$ as a function of $K_2 - K_{2c}(\chi)$.
Sufficiently far from the critical point, $\xi(\chi)$ rapidly converges to $\xi$ with increasing $\chi$.
In this regime the MPS correlation length is independent of $\chi$, and the power-law behavior of this quantity is indicative of the true critical exponent.
By comparing data for different $\chi$, we can visually determine where $\xi(\chi)$ is already sufficiently converged to the infinite-$\chi$ limit, and use only this region.
The extracted correlation length exponents on both sides of the transition are given in Fig.~\ref{fig:nu_d0.5}.
Note that the convergence of the correlation length with bond dimension is relatively gradual, thus we find a somewhat limited dynamical range of converged data $\xi(\chi \to \infty)$, presumably causing the differing values of $\nu$ on the two sides of the transition.
In addition, it is particularly evident on the VBS side that the correlation length is affected by proximity to the second-order transition to the $x$UUDD phase.
For this reason we will not use the values of $\nu$ extracted from this method in the following discussion, but rather rely on another way of determining the exponent, described below and in Fig.~\ref{fig:beta_d0.5}.

At fixed $\chi$, the MPS correlation length $\xi(\chi)$ saturates near the pseudocritical point $K_{2c}(\chi)$.
The extrapolated values from either side of the transition, denoted $\xi[\chi; K_{2c}(\chi)]$, are plotted vs $\chi$ in the insets in the corresponding panels.
Fitting to \eqref{eq:xi_vs_chi} gives similar estimates of $\kappa$ from both sides which are in rough agreement with $\kappa \approx 1.344$ expected for $c = 1$.

Considering now the order parameters, in Fig.~\ref{fig:beta_d0.5} we show $\lrangle{M_z^\text{FM}}$ (top panel) and $\lrangle{\Psi_{\VBS}}$ (bottom panel) as a function of $|K_2 - K_{2c}(\chi)|$, each within its ordered phase.
In the main plot in each, we extract the corresponding order parameter exponent over the range where we see convergence to the $\chi \to \infty$ limit.
We appear to have wider dynamical ranges for the power law fitting here compared to the correlation length data in Fig.~\ref{fig:xi_d0.5}.
The extracted order parameter exponents are roughly equal for the two order parameters, supporting one of the key predictions of the theory of the 1d DQCP.

As the order parameter scaling behavior appears to be relatively more robust compared to that of the MPS correlation length, we can try to determine the critical exponent $\nu$ via the finite-entanglement scaling of the order parameters.
Specifically, we again appeal to analogy to finite-size scaling in statistical mechanics, where in a system of length $L$ an order parameter $m$ remains finite at a critical (or pseudocritical) point and scales to zero as $L^{-\beta/\nu}$.
We conjecture that in our infinite-system MPS setup, where the bond dimension sets the cutoff length $\xi(\chi)$, the discontinuity in the order parameters at the pseudocritical point scales as
\begin{equation}
m_\text{jump} \sim \xi(\chi)^{-\beta/\nu} \sim \chi^{-\kappa\beta/\nu} ~.
\label{eq:m_vs_xi}
\end{equation}
The last expression gives the predicted scaling with bond dimension, but we will focus on $m_\text{jump}$ vs $\xi(\chi)$ which is independent of exponent $\kappa$.
For both the \zFM and VBS order parameters, the value in the optimized MPS is strictly zero on one side and non-zero on the other side of the pseudocritical point $K_{2c}(\chi)$.
Hence we obtain $m_\text{jump}$ by fitting and extrapolation of the corresponding order parameter curves $\lrangle{M_z^\text{FM}}[\chi; K_2]$ or $\lrangle{\Psi_\text{VBS}}[\chi; K_2]$ from their respective ordered sides to the pseudocritical point determined earlier.
Insets in both panels in Fig.~\ref{fig:beta_d0.5} show the corresponding $m_\text{jump}$ versus similarly obtained limiting correlation lengths at the pseudocritical points for the values of $\chi$ used in the main panels, and also show fits to the scaling form Eq.~\eqref{eq:m_vs_xi}.

The extracted values of $\beta/\nu$ are fairly close for both order parameters, in agreement with the DQCP theory prediction that $\beta_{\zFM} = \beta_{\VBS}$.
These are also roughly consistent with the estimates of $\beta$ in the main panels in Fig.~\ref{fig:beta_d0.5} and $\nu$ in Fig.~\ref{fig:nu_d0.5} made from regions where the data is converged nearest to the $\chi \to \infty$ limit, although as discussed earlier, these estimates of $\nu$ are not very accurate.
Since the extracted values of $\beta$ from the order parameter scaling appear to be more accurate than the extracted values  of $\nu$ from the correlation length scaling, we can use the estimates of $\beta$ and $\beta/\nu$ to provide a more accurate estimate of $\nu \approx 0.914 \pm 0.035$.

\subsubsection{Power law decay of correlations}
\label{subsubsec:corrs}

\begin{figure}[ht!]
\centering
\includegraphics[width=\columnwidth]{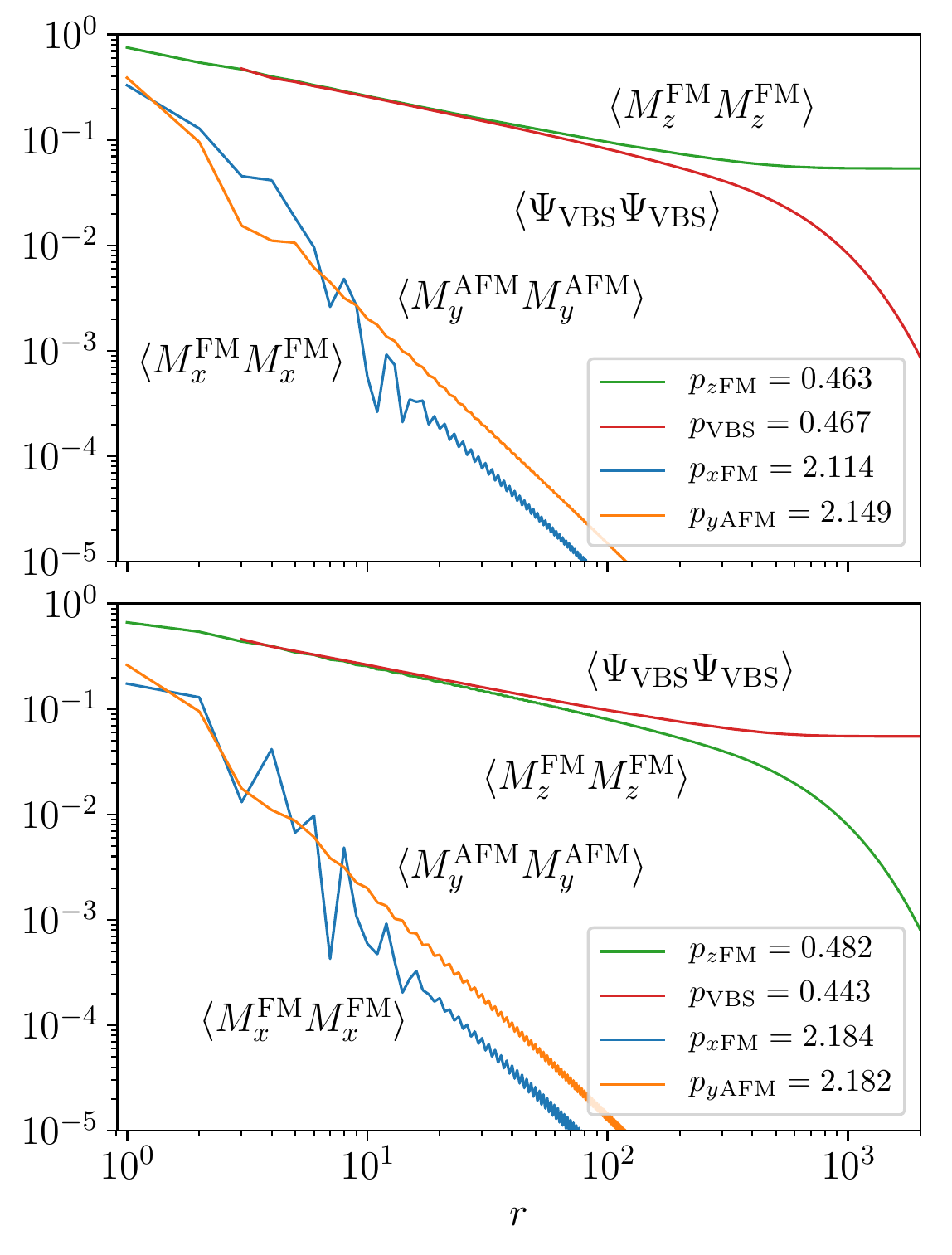}
\caption{\label{fig:corr_d0.5} (color online) Top panel shows measurements of important correlation functions (for example, $\lrangle{M_z^\text{FM} M_z^\text{FM}} \sim \lrangle{\sigma^z_0 \sigma^z_r}$) in the \zFM ansatz state tuned to the pseudocritical point $K_{2c}(\chi = 192)$.
This data is taken at $\delta = 0.5$.
All correlators show a region of critical power-law behavior before reaching a constant or decaying exponentially, as they eventually must in a finitely entangled state.
The correlation length in this state is $\sim 425$.
The bottom plot shows the same correlation functions, but measured in the \VBS ansatz MPS tuned to the pseudocritical point.
}
\end{figure}

We also measure correlation functions in our MPS in order to establish bounds on the critical decay of the important correlators in the theory introduced in Sec.~\ref{subsec:ft}.
These are $p_{\zFM}$ for $\lrangle{M_z^\text{FM} M_z^\text{FM}}$ and $p_{\VBS}$ for $\lrangle{\Psi_{\VBS} \Psi_{\VBS}}$, in addition to exponents $p_{x\text{FM}}$ and $p_{y\text{AFM}}$ for $\lrangle{M_x^\text{FM} M_x^\text{FM}}$ and $\lrangle{M_y^\text{AFM} M_y^\text{AFM}}$.
Note that the latter two correlators decay exponentially both in the \zFM and \VBS phases and only at the critical point show slower power law decay.
Examples of the correlation functions at criticality for our representative cut at $\delta = 0.5$ and the resulting bounds on the exponents are given in Fig.~\ref{fig:corr_d0.5}.

The top panel shows the correlations at the pseudocritical point $K_{2c}(\chi)$, measured in the \zFM ansatz using our largest bond dimension $\chi = 192$.
More precisely, we measure the correlations by using the adiabatic process described previously, beginning with a state well within each phase and tuning the Hamiltonian up to a very small distance $\lesssim 10^{-6}$ away from the estimated $K_{2c}(\chi)$.
In this case, the $M_z^\text{FM}$ correlations eventually saturate to a finite value while the $\Psi_{\VBS}$ correlations eventually decay exponentially (the latter is also true of the $M_x^\text{FM}$ and $M_y^\text{AFM}$ correlators).
The bottom panel shows similar measurements coming from the \VBS side, where it is now the $M_z^\text{FM}$ correlations that eventually decay exponentially while the $\Psi_{\VBS}$ correlations eventually saturate.
However, in both panels, there is a large window $r < \xi(\chi)$ where all correlators show power law decay, and we list the extracted power law exponents in each case.

Notably, we can tell even visually that the critical $M_z^\text{FM}$ correlations and $\Psi_{\VBS}$ correlations have very similar power laws, and the extracted numerical values of the exponents confirm this.
For these correlators, it is natural to take the values of the exponents extracted from the two sides as bounds on the true critical exponent; these are already fairly close, and thus provide informative bounds.
We also note that we can tell visually that the critical $M_x^\text{FM}$ and $M_y^\text{AFM}$ correlations have very close power laws; for each quantity, the extracted exponents from both sides are very close, and are also close between the two observables.

\begin{table*}[htb!]
\centering
    \begin{tabular}{c||c||c|c|c|c|c|c|c|c|c}
        $\delta$ & $K_{2 c}$ & $\beta_{\zFM}$ & $\beta_{\zFM}/\nu$ & $\beta_{\VBS}$ & $\beta_{\VBS}/\nu$ & $\nu$ & $p_{\zFM}$ & $p_{\VBS}$ & $p_{\xFM}$ & $p_{\yAFM}$ \\
        \hline
        0.1 & 0.36674 & 0.653 & 0.391 & 0.773 & 0.388 & $1.831 \pm 0.161$ & 0.759 & 0.777--0.779 & 1.290 & 1.289\\
        0.2 & 0.41087 & 0.452 & 0.279 & 0.501 & 0.361 & $1.503 \pm 0.115$ & 0.669--0.674 & 0.676--0.680 & 1.485 & 1.485\\
        0.3 & 0.45630 & 0.345 & 0.295 & 0.374 & 0.301 & $1.205 \pm 0.035$ & 0.598--0.605 & 0.593--0.602 & 1.70 & 1.71\\
        0.4 & 0.502630 & 0.269 & 0.265 & 0.282 & 0.263 & $1.045 \pm 0.029$ & 0.531--0.540 & 0.519--0.533 & 1.87 & 1.89\\ 
        0.5 & 0.549729 & 0.206 & 0.234 & 0.219 & 0.230 & $0.916 \pm 0.034$ & 0.463--0.482 & 0.443--0.467 & 2.15 & 2.17\\
        0.6 & 0.597341 & 0.156 & 0.200 & 0.167 & 0.200 & $0.808 \pm 0.027$ & 0.389--0.422 & 0.364--0.403 & 2.44 & 2.54\\ 
        0.7 & 0.644979 & 0.113 & 0.163 & 0.126 & 0.163 & $0.733 \pm 0.040$ & 0.305--0.369 & 0.287--0.347 & 3.08 & 3.15\\
    \end{tabular}
    \caption{Extracted critical points and exponents on slices of fixed $\delta$ exhibiting a direct phase transition between \zFM and \VBS.
    Quoted values for $\beta$ are extracted from near-critical scaling of the $\chi \to \infty$ converged order parameters, while values for $\beta/\nu$ are obtained from the finite-entanglement scaling of the order parameters at the precise pseudocritical points, measuring in both \zFM and \VBS ansatz states; from these, we obtain the quoted bounds on $\nu$.
    Similarly, the bounds on $p_{\zFM}$ and $p_{\VBS}$ are determined from measurements made in each ansatz state tuned to the pseudocritical point.
    }
    \label{table:exp}
\end{table*}

\subsection{Continuously varying critical exponents}
\label{subsec:varexp}

We repeat the analysis presented above in Sec.~\ref{subsec:critical_point} for multiple cuts along fixed $\delta$ which exhibit a direct \zFM to \VBS phase transition.
We conclude that this transition exists for all $\delta \leq 0.7$, with a tricritical point lying within $\delta = (0.7,0.8)$ where the transition branches, allowing an intervening phase.
We discuss this region in Sec.~\ref{sec:coex}.

Our findings for all critical exponents are summarized in Table \ref{table:exp}.
We first observe that they vary continuously with $\delta$, a general trend which is in agreement with the description of the field theory in Sec.~\ref{subsec:ft}.
Additionally, we have several specific predictions of nontrivial relationships between critical exponents which apply to any point on the phase boundary.
We test these on our cuts of constant $\delta$, finding good agreement in all cases between the predictions and observations.

\begin{figure}[ht]
\centering
\includegraphics[width=\columnwidth]{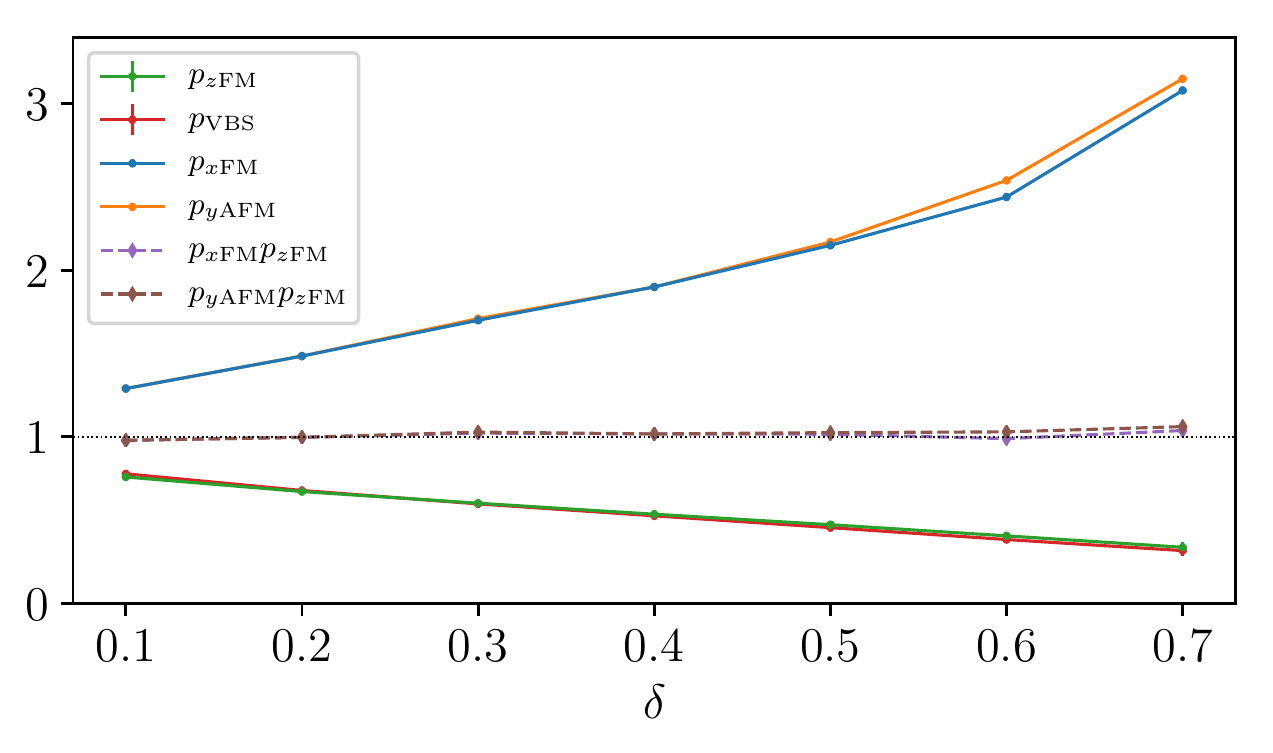}
\caption{\label{fig:cv_exp_1} (color online) The measured power law decay exponents are in good agreement with the predicted behavior $p_{\zFM} = p_{\VBS}$ and $p_{\xFM} = p_{\yAFM}$, as well as with Eq.~\eqref{eq:corr_exp}.
At the larger values of $\delta$, the state begins to feel the tricritical point, which affects the more quickly decaying $M_x^\text{FM}$ and $M_y^\text{AFM}$ correlation functions.
}
\end{figure}

Because the critical exponents in the field theory are functions of a single variable---the Luttinger parameter $\tilde g$, which varies along the critical line---they can be readily manipulated to obtain relationships between measurable quantities.
For example, we have the basic predictions that $p_{\zFM} = p_{\VBS}$ and $p_{\xFM} = p_{\yAFM}$, as well as the relationship
\begin{equation}
p_{\zFM}\; p_{\xFM} = p_{\zFM}\; p_{\yAFM} = 1~.
\label{eq:corr_exp}
\end{equation}
We find that the data are generally in good agreement with these conditions, as shown in Fig.~\ref{fig:cv_exp_1}, with some deviations for the largest $\delta$.
In this regime, the correlations feel the influence of the many other nearby critical lines, including the transitions described below which continue after the \zFM to \VBS critical line terminates.
From the power law decay exponents, we can also easily read off the Luttinger parameter: in particular, $p = \tilde{g}/2$ for the (dominant) correlations of the order parameters.
From this, we see that $\tilde g$ varies inside the expected range $(1/2,2)$.

\begin{figure}[ht]
\centering
\includegraphics[width=\columnwidth]{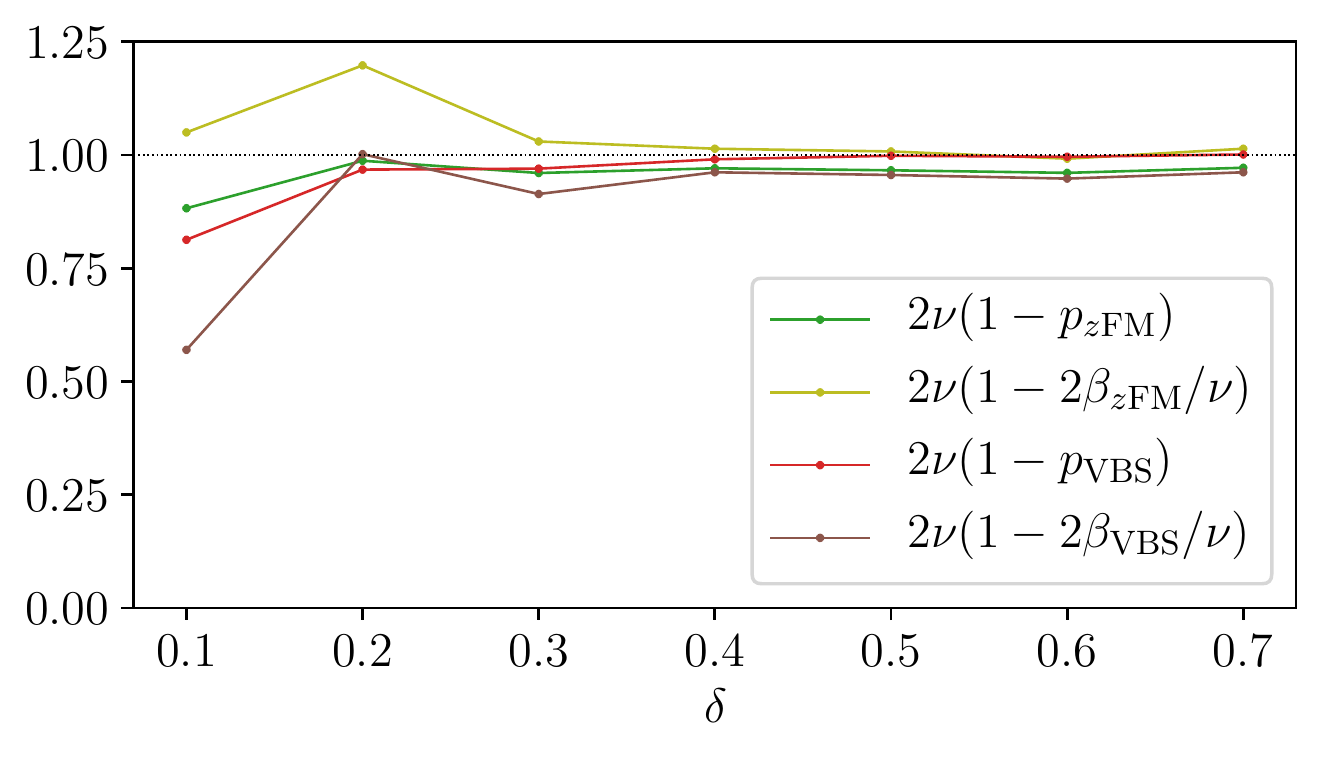}
\caption{\label{fig:cv_exp_2} (color online) We find good agreement with Eq.~\eqref{eq:nu_exp}, particularly for the larger values of $\delta$ on the critical line.
The states at small $\delta$ are near to the $\delta = 0$ QLRO phase and thus are relatively highly entangled, which makes it difficult to reach the limit $\chi \to \infty$ near enough to the critical point to extract the $\nu$ and $\beta$ critical exponents.
}
\end{figure}

From general scaling behavior we have the relationship $\beta/\nu = p/2$ for the \zFM and \VBS order parameters.
We measure both $\beta/\nu$ and $p$ directly in our MPS wavefunctions, and referring to Table \ref{table:exp} one observes that this relationship indeed holds fairly accurately.
We also have the following nontrivial prediction from the field theory:
\begin{equation}
2 \nu (1-p) = 2 \nu (1-2\beta/\nu) = 1~,
\label{eq:nu_exp}
\end{equation}
where $\beta$ and $p$ apply to the \zFM or \VBS order parameters.
We examine this prediction in Fig.~\ref{fig:cv_exp_2}, finding good agreement of the measurements with the predicted value for large $\delta > 0.3$.
However, the data at low $\delta$ exhibit some deviations from the expected behavior.
This arises from inaccuracy in our estimates of the critical exponents $\nu$ and $\beta$, which rely on convergence to the infinite-$\chi$ limit in a region near enough to the critical point to find a power-law exponent.
For low $\delta$, the state is near the quasi--long range ordered phase at $\delta = 0$ and contains a high degree of entanglement; hence, our finite-entanglement scaling is comparatively less accurate.

Despite the influence of various other nearby phases and phase transitions on our results, we have observed several nontrivial predictions from the field theory in our measurements of the continuously varying critical exponents along the \zFM to \VBS phase boundary.
This constitutes further strong evidence that this critical line is indeed an example of the DQCP described in Sec.~\ref{subsec:ft}.

\section{Study of order parameter coexistence}
\label{sec:coex}

\subsection{Evidence for coexistence regime}

Returning to the action functional in Eq.~\eqref{eq:gauged_bosonization_action}, one expects the destabilization of the \zFM to \VBS transition due to the emergence of a second relevant cosine at a critical value of the Luttinger parameter $\tilde g^\ast = 1/2$.
Here the phase transition is predicted to branch into two distinct critical lines, introducing an intermediate region where both $g_x$ and $T_1$ are broken, leading to coexistence of both order parameters.
It is not easy to relate $\tilde g$ to the microscopic parameters, but we can read off the values of the critical exponents very close to this tricritical point, finding $\nu^\ast = 2/3$, $p^\ast_{\zFM} = p^\ast_{\VBS} = 1/4$, $\beta^\ast_{\zFM} = \beta^\ast_{\VBS} = 1/12$, and $p^\ast_{\xFM} = p^\ast_{\yAFM} = 4$.

\begin{figure}[ht]
\centering
\includegraphics[width=\columnwidth]{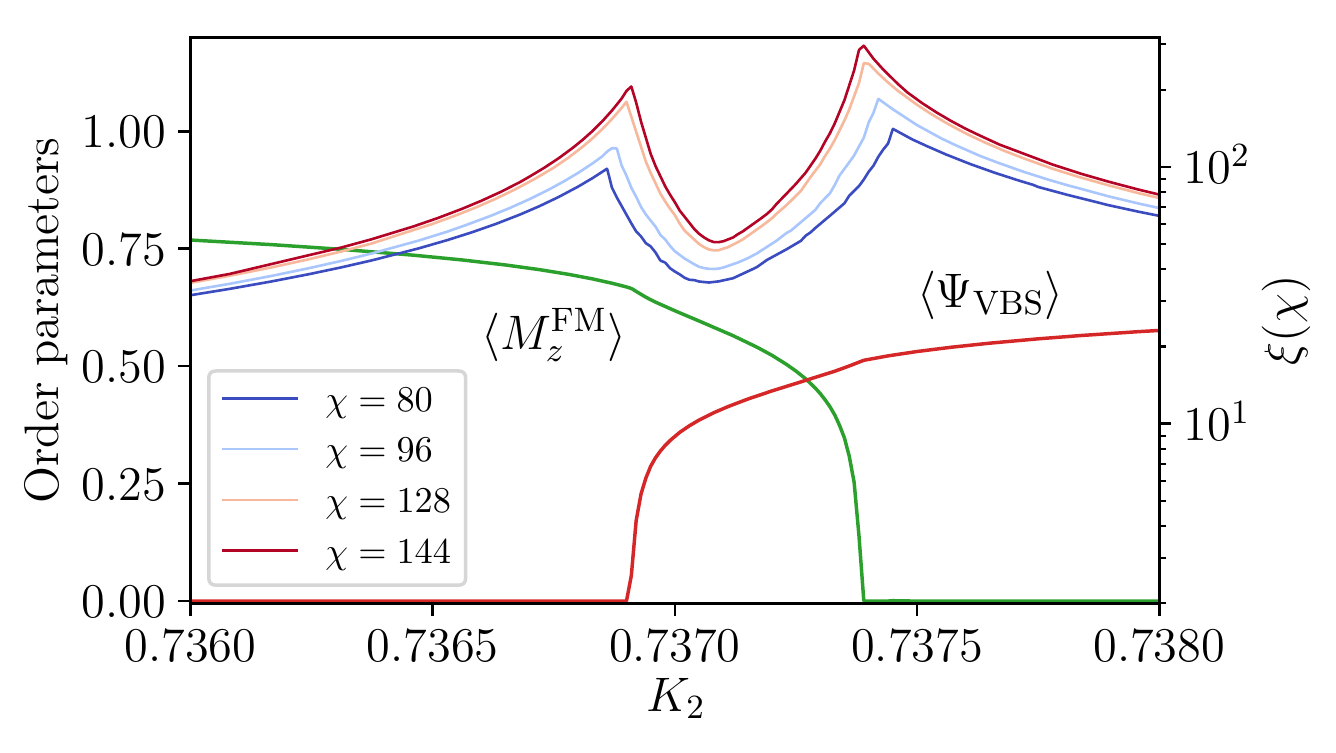}
\caption{\label{fig:op_nu_d0.9} (color online)
The slice at $\delta = 0.9$ clearly exhibits a region of coexistence of order parameters (measurements shown use the bond dimension $\chi = 144$ MPS), and the correlation length displays $\chi$-dependent cusps at both boundaries.
However, we do not have good $\chi$-converged properties inside of this phase, as the correlation length does not saturate for the bond dimensions shown here.
}
\end{figure}

We observe the branching of the phase transition at some value $\delta \in (0.7, 0.8)$, which is consistent with the description of the critical exponents given above.
The appearance of the intermediate phase is illustrated in Fig.~\ref{fig:op_nu_d0.9} for the slice $\delta = 0.9$, where the state acquires VBS order on top of the \zFM order at $K_{2c, \text{VBS}}(\chi = 144) = 0.73691$ and the \zFM order vanishes at $K_{2c, z\text{FM}}(\chi = 144)= 0.73738$.
These phase transitions are not described by the DQCP theory; rather, because in each case a single $\Z_2$ symmetry is broken, we expect the critical points to be in the Ising universality.
We explore mean field pictures of the phases in Apps.~\ref{subapp:meanfield_coexist} and \ref{subapp:coexist_mps}, finding support for this expectation.

The analysis of the boundary of the coexistence region does not follow straightforwardly from the protocol used in Sec.~\ref{subsec:critical_point}.
Because the mean field theory of these transitions is not discontinuous, we cannot exploit the level crossings of MPS trial states to accurately determine the locations of the critical points.
Similarly, we are unable to use the finite values of the order parameters at the pseudocritical points to determine critical exponents, as we do for the direct \zFM to \VBS phase transition.
In addition, as the coexistence region is very narrow and located fairly close to the $x$UUDD phase, we do not have access to a very large dynamical range.
Instead, we identify the pseudocritical points by using a power-law fit to the vanishing of the order parameters.
Also, we are able to obtain only rough estimates of the critical exponents.

We list our estimates of the transition points for $\delta = 0.8, \dots, 1.0$ in Table~\ref{tab:exp2}.
Note that the $\delta > 1$ regime can be related to $\delta < 1$ by the map described in Sec.~\ref{subsec:pt} (which related the \VBSI and \VBSII phases), so numerical studies are required only for $\delta \leq 1$.
Also, while for $\delta \neq 1$ the \zFM-ordering transition $K_{2c, z\text{FM}}$ involves a strictly \VBS ordered phase, for $\delta = 1$ the situation is more complex and the coexistence phase actually transitions to the $x$UUDD phase.
(See the inset in Fig.~\ref{fig:phase_diagram} for an image of the coexistence region.)
We first focus on $\delta < 1$ and consider the $\delta = 1$ case later.

\begin{figure}[ht]
\centering
\includegraphics[width=\columnwidth]{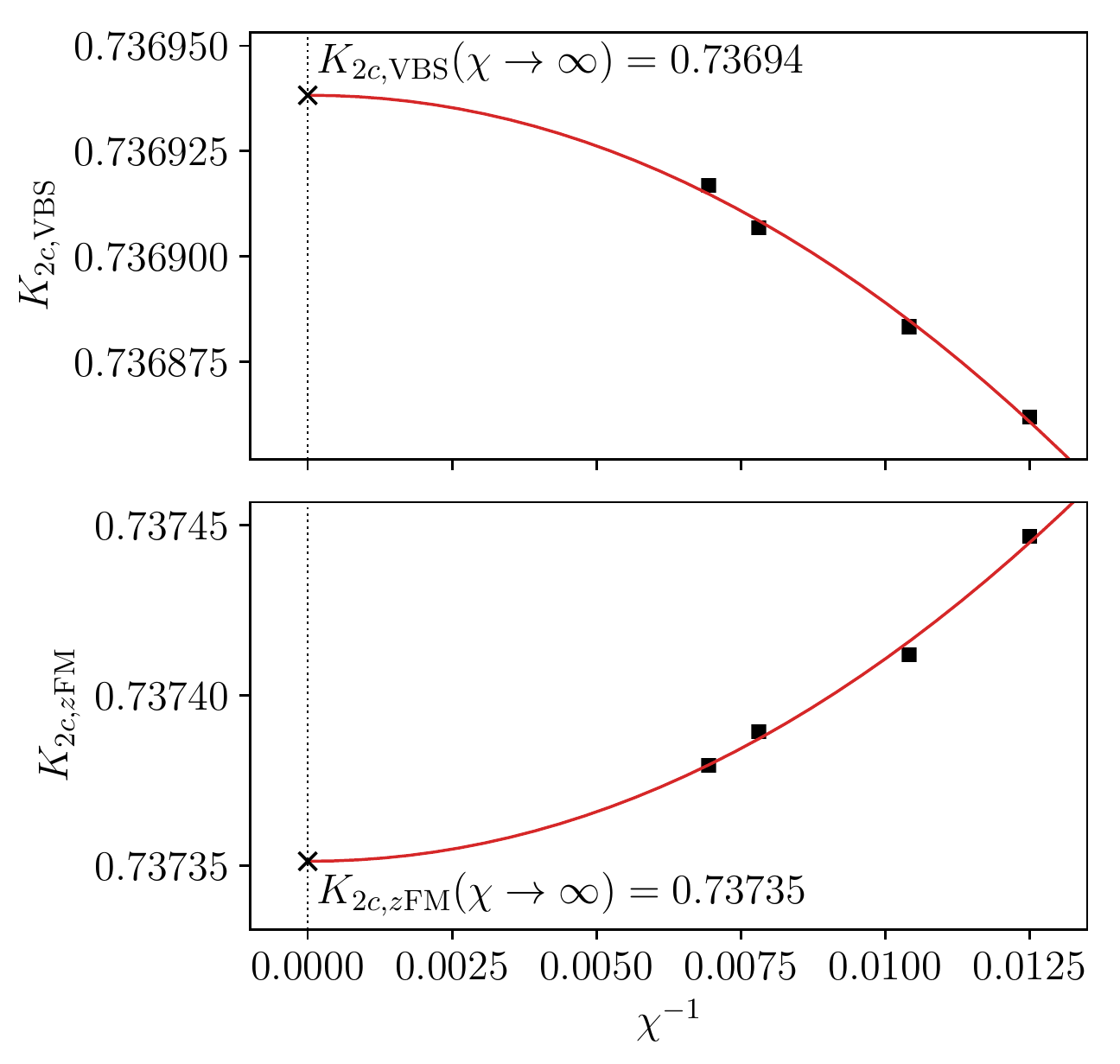}
\caption{\label{fig:K2c_d0.9} (color online)
Illustration of the process of locating the critical point for each order parameter from finite-entanglement scaling in the coexistence regime at $\delta = 0.9$.
Here the data points are found via fits to the power law onset behavior shown in Fig.~\ref{fig:op_nu_d0.9}.
Again, using the finite-entanglement scaling form Eq.~\eqref{eq:K2c_scal}, we extrapolate from  pseudocritical $K_{2c,\VBS}(\chi)$ and $K_{2c,\zFM}(\chi)$ to estimate the width of the coexistence region in the limit $\chi \to \infty$.}
\end{figure}

\begin{figure}[ht]
\centering
\includegraphics[width=\columnwidth]{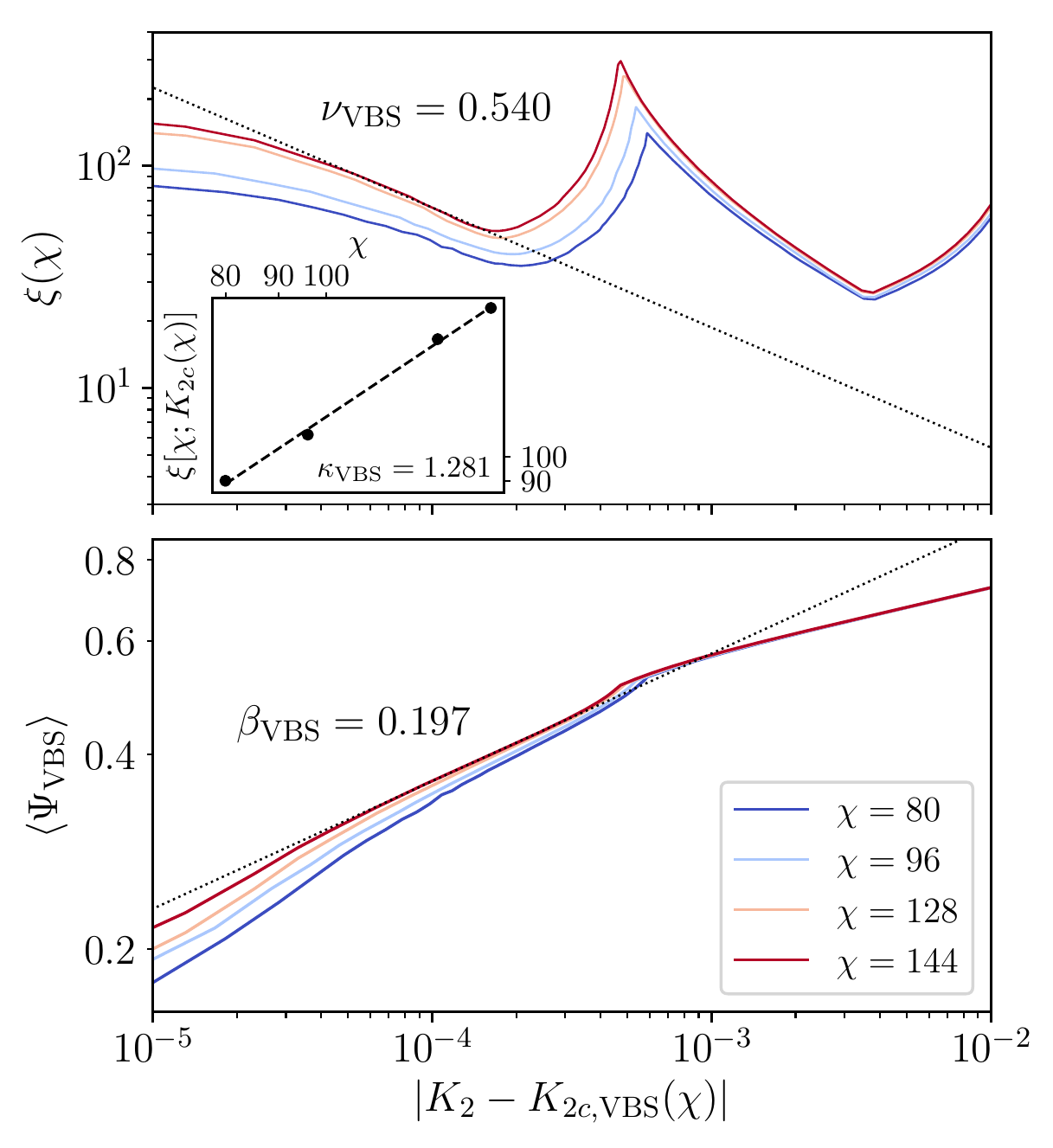}
\caption{\label{fig:vbs_pt_d0.9} (color online) Our study of the \VBS ordering transition in the coexistence region is impeded by the narrow width of the phase, here shown at $\delta=0.9$.
The top panel shows correlation length along with our best fit, though the MPS results are not reflective of the $\chi \to \infty$ limit and the exponent is far from the Ising $\nu = 1$.
The feature seen near $3\times 10^{-4}$ on the $x$-axis is the \zFM order transition on the boundary of the coexistence region with the \VBSI phase (this transition is studied in Fig.~\ref{fig:zfm_pt_d0.9}).
Further from the critical point, one sees the effect of the transition to the $x$UUDD phase.
The bottom panel shows the onset of the \VBS order parameter, which is roughly consistent with a continuous phase transition but does not agree with the Ising $\beta = 1/8$. 
Data here do not use the adiabatic protocol; every point is independent.
}
\end{figure}

\begin{figure}[ht]
\centering
\includegraphics[width=\columnwidth]{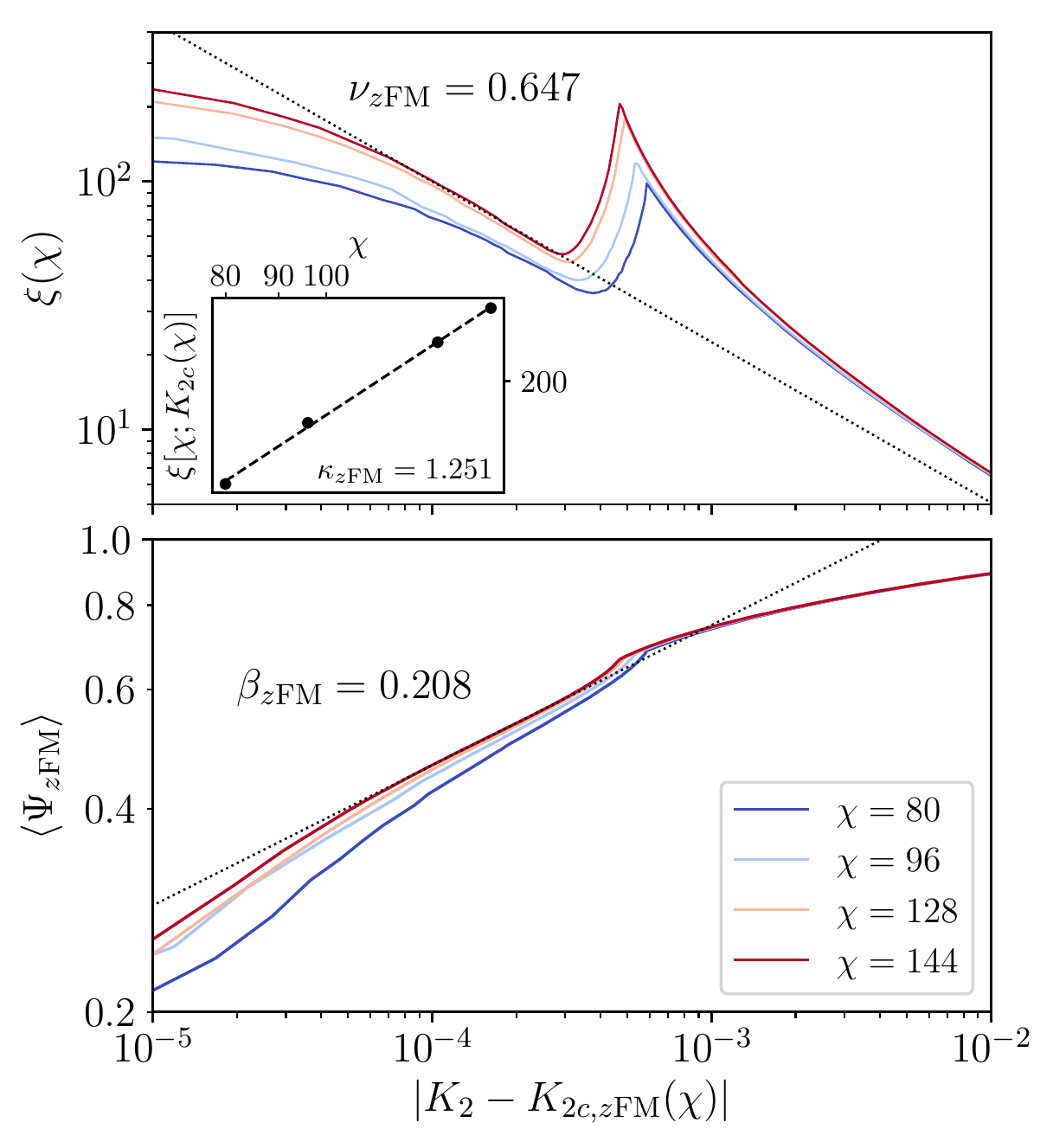}
\caption{\label{fig:zfm_pt_d0.9} (color online) Similarly to Fig.~\ref{fig:vbs_pt_d0.9}, we provide only a rough study of the \zFM ordering transition in the coexistence region, here shown at $\delta=0.9$.
The top panel shows correlation length along with our best fit, though the MPS results are not reflective of the $\chi \to \infty$ limit and the exponent is far from the Ising $\nu=1$.
The feature seen near \mbox{$3\times 10^{-4}$} on the $x$-axis is the \VBS order transition on the boundary of the coexistence region with the \zFM phase (this transition is studied in Fig.~\ref{fig:vbs_pt_d0.9}).
The bottom panel shows the onset of the \zFM order parameter, which is roughly consistent with a continuous phase transition but does not agree with the Ising $\beta = 1/8$. 
Data here do not use the adiabatic protocol; every point is independent.
}
\end{figure}

Figure~\ref{fig:op_nu_d0.9} shows the correlation length and expectation values of the order parameters in the coexistence region.
We observe cusps in the correlation length at both transitions, with height which increases with increasing $\chi$.
However, our data are not of sufficient granularity to perform definitive finite-entanglement scaling at these critical points, and we do not have sufficient dynamical range between the two critical points to extract the correlation length exponent at either transition.
On the \VBS side, we are also close to the $x$UUDD phase boundary.
We can attempt to find $\nu$ looking at the pure \zFM side; however, the width of the crossover region where the correlation length saturates in $\chi$ is significantly wider than the distance between $K_{2c,\VBS}$ and $K_{2c,\zFM}$.
In this case, the extracted $\nu$ likely does not cleanly correspond to just one transition but instead combines information about all nearby phase transitions and even the tricritical point.
Thus, in the top panels of Figs.~\ref{fig:vbs_pt_d0.9} and \ref{fig:zfm_pt_d0.9}, we focus only on the data from the coexistence region, with the understanding that they will hardly be conclusive.
The extracted values of $\kappa$ are quite far from the expectation for this $c=1/2$ critical point.
In the bottom panels of these figures, we have attempted to extract the order parameter onset exponents at each critical point.
By the same argument, we clearly should restrict attempts at fitting power-law onset forms to be within the coexistence region.
However, we see that the apparent slopes continue to vary visibly for our range of bond dimensions $\chi$.
In particular, these measurements are likely to be influenced by some mixture of the actual Ising criticality as well as the mean-field phase transition in the MPS at the pseudocritical point, and indeed we find values for the critical exponents that lie between these two.

\begin{table*}[ht]
    \centering
    \begin{tabular}{c || c | c | c | c || c | c | c | c}
        $\delta$ 
        & $K_{2c,\VBS}$ & $\nu_{\VBS}$ & $\beta_{\VBS}$ & $p_{\VBS}$
        & $K_{2c,\zFM}$ & $\nu_{\zFM}$ & $\beta_{\zFM}$ & $p_{\zFM}$ \\
        \hline
        0.8 & 0.691922 & $\cdot$ & $\cdot$ & 0.29 & 0.691927 & $\cdot$ & $\cdot$ & 0.30 \\
        0.85 & 0.71481 & $\cdot$ & $\cdot$ & 0.30 & 0.71486 & $\cdot$ & $\cdot$ & 0.28 \\
        0.9 & 0.73693 & 0.54 & 0.20 & 0.33 & 0.73735 & 0.65 & 0.21 & 0.26 \\
        0.95 & 0.75798 & 0.63 & 0.19 & 0.40 & 0.75936 & 0.68 & 0.20 & 0.28
    \end{tabular}
    \caption{Critical properties at the \VBS ordering transition $K_{2c,\VBS}$ between the \zFM and coexistence phases, and at the \zFM ordering transition $K_{2c,\zFM}$, between the \VBS and coexistence phases.
    All data is measured within the coexistence region, in order to reduce the effects of other nearby criticalities.
    The transition is too narrow for $\delta < 0.9$ to allow for the determination of the correlation length and order parameter onset critical exponents.}
    \label{tab:exp2}
\end{table*}

Table~\ref{tab:exp2} summarizes our estimates of the critical indices for the transitions on the slices $\delta=0.9$ and $0.95$.
These are rather inaccurate, as explained above, and are shown to emphasize our limitations when studying the transitions involving the coexistence phase.
We also quote estimates of the power law correlation decay exponents extracted from fits at the corresponding pseudocritical points for our largest $\chi = 144$.
These estimates also differ somewhat from the exponent $p=1/4$ expected at each Ising transition, but the accuracy may be a bit better than for the extracted $\nu$ and $\beta$ values.

\subsection{Higher-symmetry line at \texorpdfstring{$\delta = 1$}{delta = 1}}
\label{sec:delta1}

\begin{figure}[ht]
\centering
\includegraphics[width=\columnwidth]{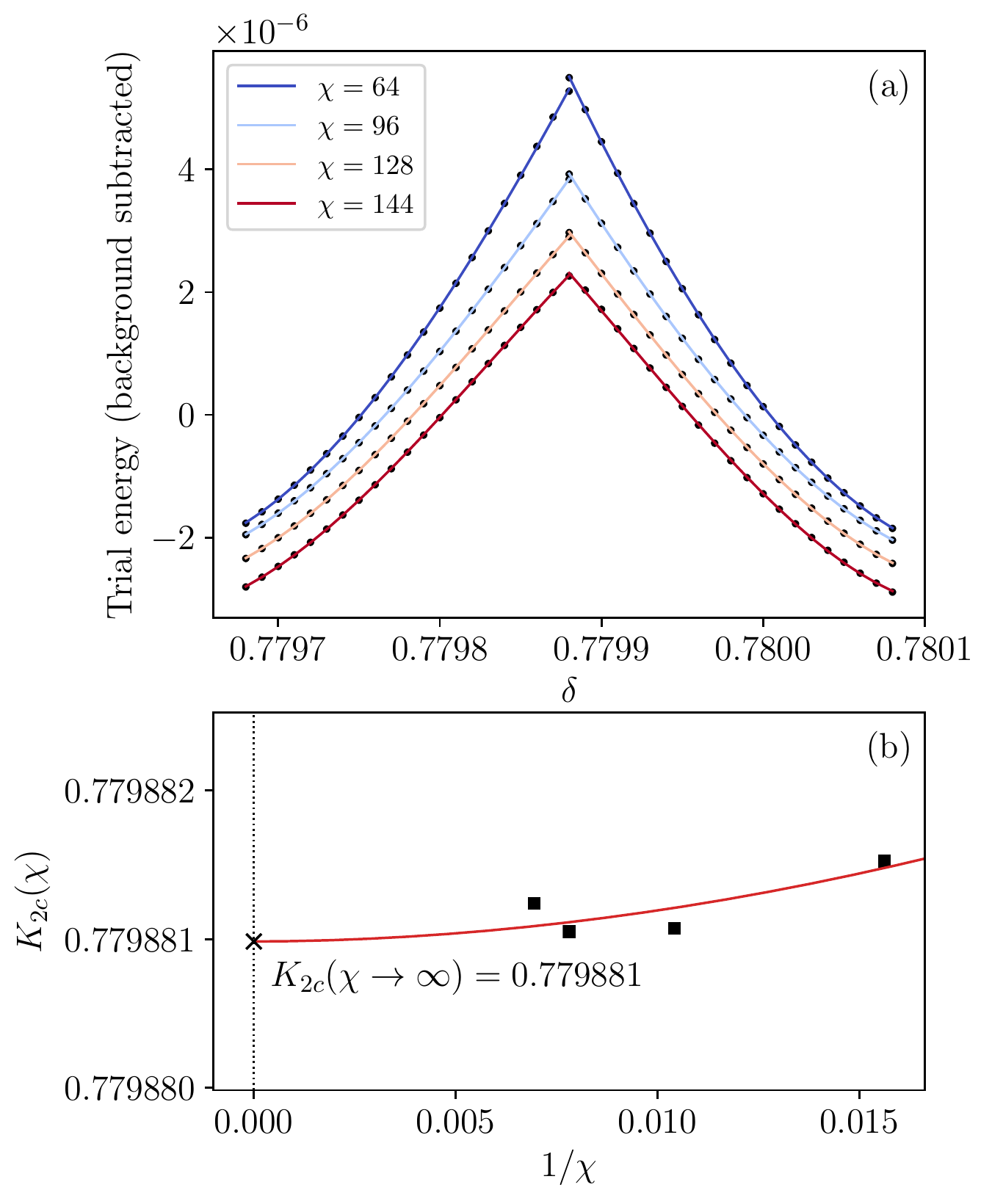}
\caption{\label{fig:K2c_d1.0} (color online)
Illustration of determining the critical point between \zFM+\VBS coexistence and $x$UUDD from finite-entanglement scaling at $\delta = 1$, performed using the same procedure used in Fig.~\ref{fig:K2c_d0.5}.
(a) The energies of both the coexistence and $x$UUDD phases follow smooth curves which determine the level crossing, but here happen to not display hysteresis.
This does not present a problem, as the smoothness of the evolution of the trial state energies permits extrapolation.
(b) Using a fit to the finite-entanglement scaling form Eq.~\eqref{eq:K2c_scal}, we extrapolate the pseudocritical $K_{2c,\text{zFM}}(\chi)$ to estimate the location of the critical point at $\chi \to \infty$.
As was the case for the DQCP, the scatter in data points is again not noise from the variational algorithm, but a reproducible feature of the ground state at each bond dimension.}
\end{figure}

\begin{figure}[ht]
\centering
\includegraphics[width=\columnwidth]{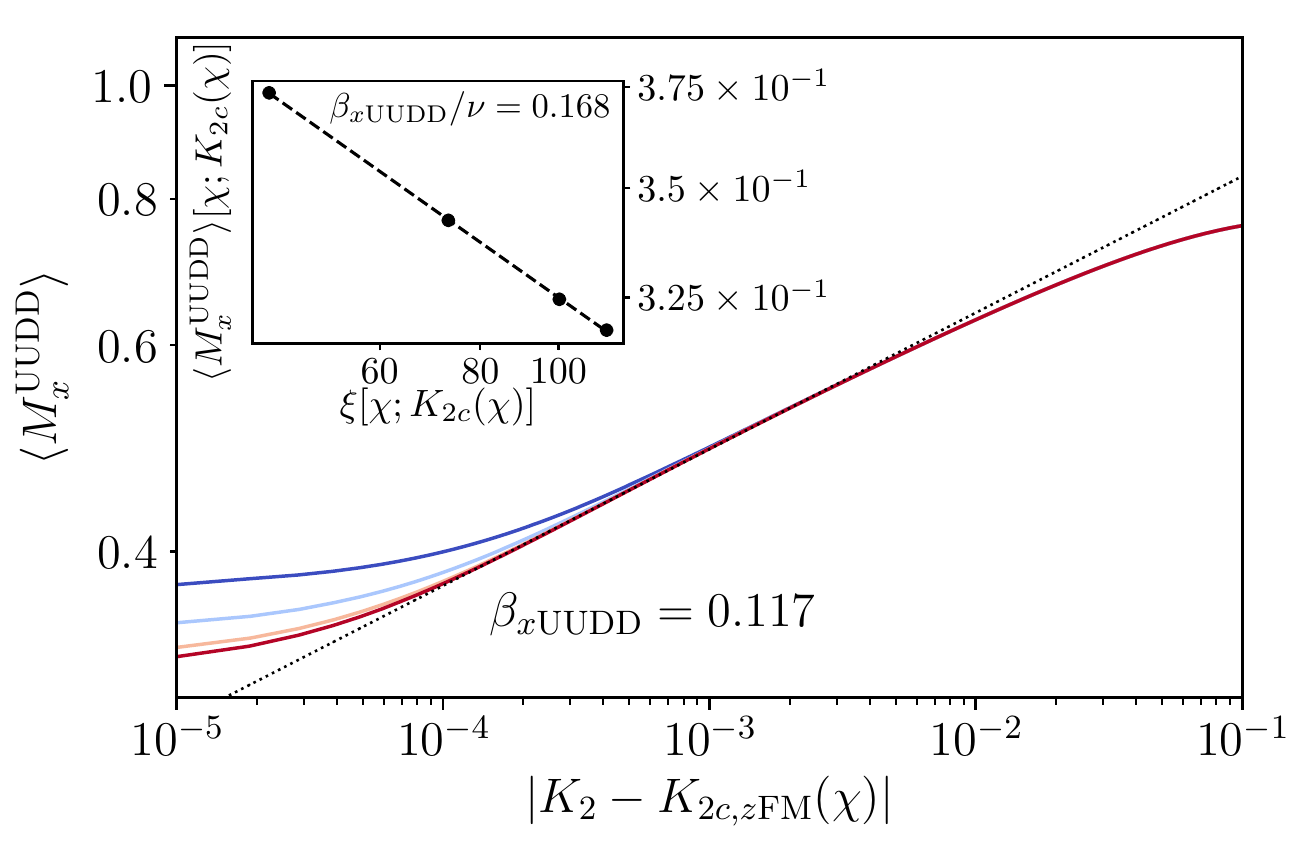}
\caption{\label{fig:beta_d1.0} (color online)
Analysis of the phase transition from the \zFM+\VBS coexistence phase to the $x$UUDD phase along the cut $\delta=1$, using the discontinuous VUMPS (generalized mean field) procedure.
Because we cannot make an accurate determination of $\beta_{\zFM}$ using measurements arising from inside the coexistence phase, we consider only measurements of $M_x^\text{UUDD}$, within the $x$UUDD phase.
However we can perform finite-entanglement scaling analysis of $M_z^\text{FM}$ extrapolated to the pseudocritical points.
Doing so, we find a value $\beta_{\zFM}/\nu = 0.167$ which is quite similar to $\beta_{x\text{UUDD}}/\nu$ shown in the inset.
}
\end{figure}

The line $\delta = 1$ admits an additional symmetry of the Hamiltonian:
\begin{equation}
g_{z, \text{even}} = \prod_m \sigma^z_{2m}~: \quad \sigma^z_{2m} \mapsto \sigma^z_{2m},\;\sigma^{x,y}_{2m} \mapsto -\sigma^{x,y}_{2m}~.
\end{equation}
This is the same symmetry which takes \VBSI to \VBSII, and vice versa.
As we stated previously, one possibility for these two phases along this cut is a first-order transition, but this turns out not to be the case in our model.
Instead, inside of the coexistence region where $g_x$ is broken the \VBSI and \VBSII orders are in fact the same.
This was pointed out in Ref.~\cite{jiang19}, and we provide a demonstration in Apps.~\ref{subapp:meanfield_coexist} and \ref{subapp:coexist_mps} by writing fixed-point wavefunctions for the coexistence which smoothly interpolate between \VBSI and \VBSII in the presence of $g_x$ symmetry breaking.
The onset of the \VBS order---the boundary between the \zFM and coexistence phases---is thus no different from the case for other $\delta$.

On the other hand, at the \zFM ordering transition (that is, the transition out of the coexistence phase at which the \zFM order disappears), the $g_x$ symmetry is restored.
Here we find a transition not to a state with pure \VBS character, but rather to the $x$UUDD phase.
Because the phases on either side break different $\Z_2$ symmetries yet we observe a direct phase transition, this criticality in fact bears a resemblance to the \zFM to \VBS DQCP studied in the preceding sections.
(Here the direct transition between the coexistence and $x$UUDD phases is enforced by the additional $\Z_2$ symmetry $g_{z, \text{even}}$ which apparently remains unbroken in our model.)
In fact, in App.~\ref{app:delta1} we develop a theory of this transition which turns out to be similar to the critical line at $\delta = 0$ separating the \zFM and \xFM phases but placed on top of a translation symmetry--breaking background.

We are able to study this transition using the methods of Sec.~\ref{sec:mps_study}, where now instead of the \VBS order parameter $\Psi_{\VBS}$ (which remains ordered throughout the transition) we measure
\begin{equation}
\lrangle{M_x^\text{UUDD}} = \lrangle{\sigma^x_0}~,
\end{equation}
where, as was the case for the previous order parameters, there is a sign ambiguity which we ignore.
In a more precise sense, the order parameter in the $x$UUDD phase has two Ising-like components, $\left(\sum_j (-1)^{j(j-1)/2} \sigma^x_j,~ \sum_j (-1)^{j(j+1)/2} \sigma^x_j \right)$, where in each case $j$ runs over the unit cell.
However, because the MPS ground state always has spontaneously broken symmetry---that is, the ``UUDD'' pattern or its partners related by translation---it suffices to confirm this pattern and measure just $\lrangle{\sigma^x_0}$.
Because of the relatively slow convergence in $\xi$ exhibited by the correlation length in Sec.~\ref{subsubsec:nu_beta_d0.5} as well as the limited dynamical range within the coexistence region, we focus only on measurements of the order parameter in the $x$UUDD phase to characterize this transition.
This study is shown in Fig.~\ref{fig:beta_d1.0}, where we find that $\beta_{x\text{UUDD}} = 0.117$ and $\nu = 0.69$.
We also find from measurement of the power-law decaying correlation functions that $p_{\xFM} \approx 0.39$ and $p_{\zFM} \approx 0.37$.
These values provide some point of reference relative to the other phase transitions studied in this work but are not significant by themselves, as this critical point lies on a line exhibiting continuously varying exponents.
Because this line crosses our phase diagram plane only at one point, in the present study we cannot observe continuously varying exponents but we do check that the expected relationships are approximately satisfied: $p_{\xFM} \approx p_{\zFM}$ and $2 \nu (1-p_{\xFM}) \approx 0.86$ (compared to the expected value 1).

\section{Conclusion}
\label{sec:conclusion}

We performed a detailed numerical study of the ferromagnet to VBS transition in a spin-1/2 chain with $\Z_2 \times \Z_2$ symmetry and confirmed key predictions of the 1d DQCP theory of Ref.~\onlinecite{jiang19}.
Namely, the \zFM and \VBS order parameters have equal scaling dimensions, and the $x$FM and $y$AFM correlations of secondary importance also have equal power law exponents at the \zFM to \VBSI transition (the fact that the next-most important observables are the ferromagnetic component of $\sigma^x$ and the antiferromagnetic component of $\sigma^y$ is related to the 
crystalline SPT--like property of the \VBSI phase that distinguishes it from the \VBSII phase, which is also realized in our model).
All exponents vary continuously along the phase boundary but are controlled by a single parameter; this implies relationships among the various exponents, which we confirmed in our numerics.
The observed range of the variation of the critical indices is consistent with the regime of validity of the proposed field theory, and we also found the predicted splitting of the transition and appearance of the \VBS+\zFM coexistence phase at one end of this range.
Interestingly, we also found an instance of a new Landau-forbidden transition between the \VBS+\zFM and $x$UUDD phases along the line $\delta = 1$ with the additional $\Z_2$ symmetry $g_{z,\text{even}}$.

In our study of the 1d DQCP, we found that VUMPS at fixed bond dimension shows a discontinuous transition at a $\chi$-dependent pseudocritical point, and argued that this is related to the non-Landau nature of the transition which gives first-order behavior in the mean field.
We used this discontinuous nature to our advantage to find the pseudocritical points very accurately and 
for subsequent ``finite-entanglement'' scaling.
We propose that this protocol can be very useful at all transitions described by DQCP, and indeed we have already used it at the new direct continuous \VBS+\zFM to $x$UUDD transition enforced by the additional $g_{z, \text{even}}$ symmetry.
To accurately locate pseudocritical points is more difficult at conventional continuous transitions where the mean field is also continuous~\cite{liu10}, but it can be a powerful systematic approach in such cases as well.

In this work, we focused on static (i.e., equal time) properties at the transition.
It would be interesting to also study dynamical properties at the transition, both numerically and analytically, to see if they reveal more signatures of fractionalized excitations at this DQCP, in the spirit of the 2d study in Ref.~\onlinecite{MaSunYouXuVishwanathSandvikMeng2018}.
Analytically, we can calculate dynamical structure factors at low frequencies using the effective field theory description, but we can also try to capture properties at high frequencies and high momenta using one of the microscopic parton descriptions in Ref.~\onlinecite{jiang19}, for example using the fermionic parton mean field.

Finally, it would also be interesting to search for other instances of DQCP in 1d, in particular with more complex symmetries that can occur for higher spin or $\mathbb{Z}_N$ clock degrees of freedom.
Both numerical and analytical studies are needed, and one guide here is to look for systems with an LSM-type theorem.

\begin{acknowledgments}

The authors would like to thank A.~L{\"a}uchli, C.-J.~Lin, Y.-M.~Lu, Y.~Ran, A.~Sandvik, A.~Vishwanath, C.~White, and Y.-Z.~You for useful discussions.
This work was supported by the Institute for Quantum Information and Matter, an NSF Physics Frontiers Center, with support of the Gordon and Betty Moore Foundation, and also by the NSF through grant DMR-1619696.

{\it Note added:} We would like to draw the reader's attention to a related parallel work by Rui-Zhen Huang, Da-Chuan Lu, Yi-Zhuang You, Zi Yang Meng, and Tao Xiang, to appear in the same arXiv posting.

\end{acknowledgments}

\appendix

\section{Mean-field study of phase diagram with separable states}
\label{app:meanfield_sep}

In this appendix, we present caricature (``fixed-point") wavefunctions for the phases of interest in our model and use these as simple trial states to find a mean-field phase diagram of the model.
Besides developing basic intuition about the phases and their competing energetics, we demonstrate that in the mean field treatment the \zFM to \VBS transition is first-order, while the \VBS to \zFM+\VBS and also the \VBS to $x$UUDD transitions are second-order.
This provides some understanding of the observed ``pseudocritical'' behavior of VUMPS at fixed bond dimension $\chi$, i.e., behavior very close to $K_{2c}(\chi)$.

\subsection{Trial states without variational parameters}
\label{subapp:primitive_trial_states}

The \zFM fixed-point state is simply
\begin{equation}
\ket{z\text{FM}} = \otimes_j \, \ket{\up}_j
\end{equation}
or its counterpart $g_x \ket{z\text{FM}}$, with average energy per site
\begin{equation}
\epsilon_{z\text{FM}} = -J_z + K_{2z} = -(1 + \delta) + K_2 ~.
\label{eq:epsilon_zFM}
\end{equation}
In the right-hand side above, as well as in other trial energy expressions below, we specialize to the slice in the parameter space used in the main text, namely $J_z = 1 + \delta$, $J_x = 1 - \delta$, $K_{2z} = K_{2x} = K_2$.
Note that this wavefunction is an exact ground state at $\delta = 1$, $K_2 = 0$.

The \VBSI fixed-point state is
\begin{equation}
\ket{\VBSI} = \otimes_m \ket{D_{2m-1, 2m}^{(I)}}
\end{equation}
or its counterpart $T_1 \ket{\VBSI}$, where the elementary dimer state of two spins is given in Eq.~\eqref{eq:d12-I}.
The average energy per site is
\begin{equation}
\epsilon_{\VBSI} = -(J_z + J_x)/2 = -1 ~.
\end{equation}
This wavefunction is an exact ground state at the Majumdar--Ghosh point $\delta = 0$, $K_2 = 0.5$~\cite{MajumdarGhosh1969, MajumdarGhosh1969II, FurukawaSatoFurusaki2010, FurukawaSatoOnodaFurusaki2012}.

The \VBSII fixed point state is
\begin{equation}
\ket{\VBSII} = \otimes_m \ket{D_{2m-1, 2m}^{(II)}}
\end{equation}
or its counterpart $T_1 \ket{\VBSII}$, where the corresponding dimer state of two spins is given in Eq.~\eqref{eq:d12-II}.
The average energy per site is
\begin{equation}
\epsilon_{\VBSII} = -(J_z - J_x)/2 = -\delta ~.
\end{equation}
This wavefunction becomes an exact ground state for \mbox{$\delta \to \infty$,} $K_2/\delta = 0.5$.

The $x$UUDD fixed-point state is
\begin{equation}
\ket{x\text{UUDD}} = \otimes_n \ket{+\hat x}_{4n-3}\, \ket{+\hat x}_{4n-2} \, \ket{-\hat x}_{4n-1} \, \ket{-\hat x}_{4n} 
\label{eq:psi_xUUDD}
\end{equation}
along with its symmetry counterparts $T_1 \ket{x\text{UUDD}}$, $(T_1)^2 \ket{x\text{UUDD}} = g_z \ket{x\text{UUDD}}$, $(T_1)^3 \ket{x\text{UUDD}}$.
The average energy per site is
\begin{equation}
\epsilon_{x\text{UUDD}} = -K_{2x} = -K_2 ~.
\end{equation}
This wavefunction is an exact ground state for the general model at $J_z = 0$, $K_{2z} = 0$, $K_{2x} > J_x/2$, while it does not occur as a ground state on our slice through the parameter space with $K_{2z} = K_{2x}$.
Note that our definition of this phase is that it breaks the $g_z$ and $T_1$ symmetries but preserves $g_x$ and $g_z (T_1)^2$; hence, the ground state degeneracy is four.
The above wavefunction is the only product state that satisfies these symmetries.
The above ground state manifold has an additional symmetry $T_1 g_{z,\text{even}}$, which is not a symmetry of the Hamiltonian and is hence spurious, except at $\delta = 1$; in App.~\ref{subapp:improved_xUUDD} below we write improved variational wavefunctions without this spurious symmetry.

\begin{figure}
\centering
\includegraphics[width=\columnwidth]{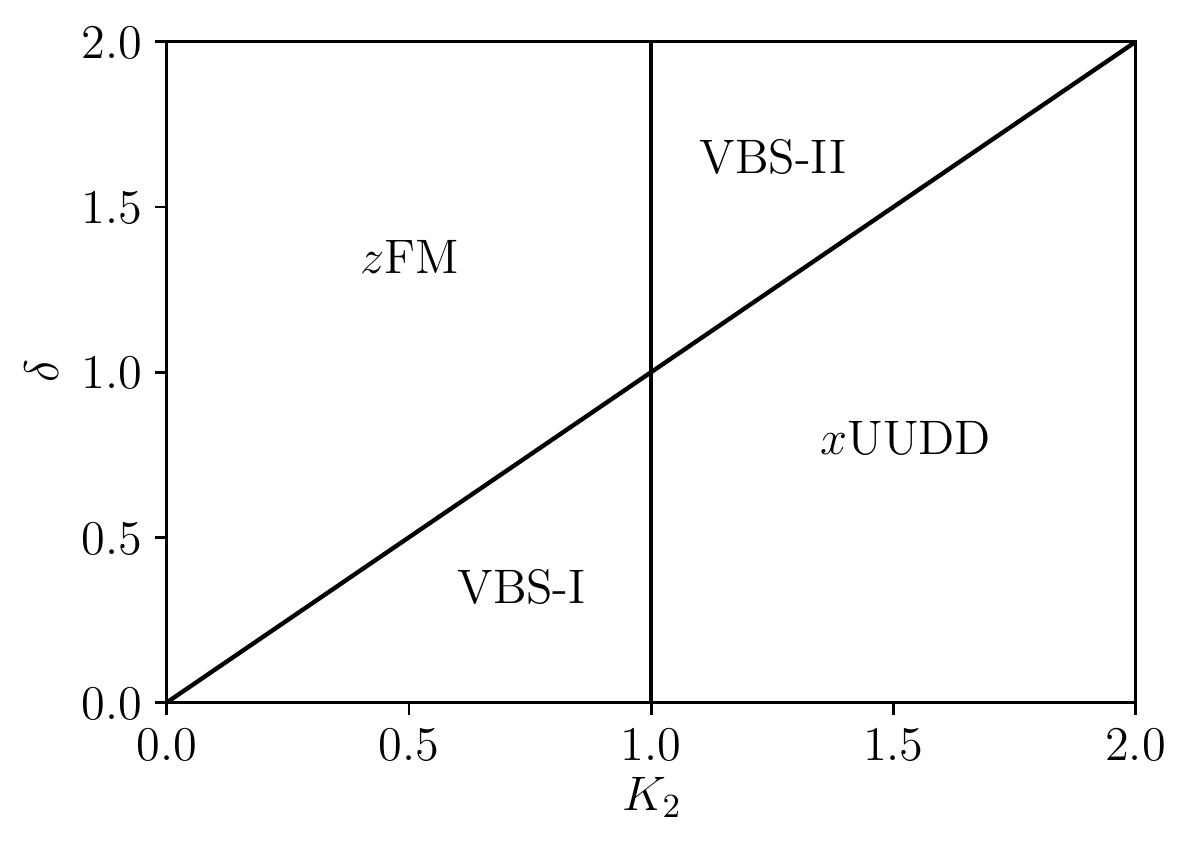}
\caption{\label{fig:mfphased} Comparing the energies of the separable trial wavefunctions of App.~\ref{subapp:primitive_trial_states} results in a phase diagram which is broadly similar to the actual behavior of the model, but renders all phase transitions as first-order.
}
\end{figure}

Comparing the trial energies $\epsilon_{\zFM}$, $\epsilon_{\VBSI}$, $\epsilon_{\VBSII}$, and $\epsilon_{x\text{UUDD}}$, we obtain the mean field phase diagram in Fig.~\ref{fig:mfphased}.
All solid lines in this figure represent ``level crossings'' and are first-order phase boundaries.
The positioning of the phases is roughly similar to the actual phase diagram in the main text, but, of course, this simple mean field is not quantitatively accurate and fails qualitatively about the nature of the \zFM to VBS transition.

\subsection{Dimer product states for the \texorpdfstring{\zFM}{zFM} and \texorpdfstring{\VBS}{VBS} coexistence}
\label{subapp:meanfield_coexist}

We can also allow for coexistence between the \zFM and \VBS order parameters, for example by using a trial state of the form
\begin{equation}
\ket{\VBS+\zFM} = \otimes_m
\left[\cos\frac{\alpha}{2}\; \ket{\up\up} + \sin\frac{\alpha}{2}\; \ket{\dn\dn} \right]_{2m-1, 2m} ~.
\label{eq:psi_VBSzFM}
\end{equation}
Clearly, at $\alpha = \pi/2$ and $-\pi/2$ the wavefunction reduces to $\ket{\VBSI}$ and $\ket{\VBSII}$ respectively, and $\alpha = 0$ gives $\ket{\zFM}$; for generic $\alpha$ the state has both \VBS and \zFM order.
The trial energy per site is
\begin{align}
\epsilon_{\VBS+\zFM} &= \frac{-J_z (1 + \cos^2\alpha) - J_x \sin\alpha}{2} + K_{2z} \cos^2\alpha \nonumber \\
&= -J_z + K_{2z} - \frac{J_x}{2}t +\left(\frac{J_z}{2} - K_{2z}\right) t^2 ~,
\end{align}
where $t \equiv \sin\alpha$, $|t| \leq 1$.
For $K_{2z} > J_z/2 - |J_x|/4$, the lowest energy is achieved at $t = \text{sign}(J_x)$, which corresponds to pure \VBSI or \VBSII order.
Thus, large $K_2$ prefers the pure dimer states.

Conversely, for $K_{2z} < J_z/2 - |J_x|/4$, this mean field finds it favorable to have coexistence of the \VBS and \zFM orders, with the optimal $t = J_x/[2(J_z - 2K_{2z})]$ and the trial energy $\epsilon_{\VBS+\zFM} = -J_z + K_{2z} - J_x^2/[8 (J_z - 2K_{2z})]$ that is always lower than the product state \zFM trial energy Eq.~\eqref{eq:epsilon_zFM} except at $J_x = 0$.
(See Fig.~\ref{fig:mfphased2}, which also includes competition with improved $x$UUDD states.)
We know that this feature is not found in our model beyond mean field, where in fact it is the pure \zFM phase that wins over the coexistence phase for small $K_2$.
This artifact arises from the fact that for the pure \zFM phase we used a trial state with zero entanglement, whereas for the coexistence phase we allowed entanglement on alternating bonds, which apparently always lowers the energy.
This lowering of the energy while simultaneously breaking the translation symmetry is undesirable in the true ground state for small $K_2$:
For example, for $K_2 = 0$ and small $J_x$ the second-order perturbation theory on top of the fixed-point \zFM state lowers the energy by $-J_x^2/(4J_z)$ per site---which is better than $\epsilon_{\VBS+\zFM}$---but to capture this lowering one needs to allow entanglement on all bonds.

On the large $K_2$ side, the mean field transition between either of the \VBS phases and \VBS+\zFM is continuous.
We thus expect that numerics at fixed bond dimension $\chi$ will show a continuous mean field--like transition at the pseudocritical point $K_{2c,\zFM}(\chi)$, which is indeed what we observe and use to locate $K_{2c,\zFM}(\chi)$ and extrapolate to the true $K_{2c,\zFM}(\chi \to \infty)$.
Of course, the true \VBS to \VBS+\zFM transition is characterized by the onset of the \zFM order on top of ``inert'' background \VBS order and is expected to be in the Ising universality class.
We also expect that the true \zFM to \VBS+\zFM transition is in the Ising universality.
While our primitive mean field does not realize this transition, we expect that VUMPS using fixed $\chi$ will have a continuous mean field--like \zFM to \VBS+\zFM transition at the corresponding pseudocritical $K_{2c,\VBS}(\chi)$, which is again borne out in the numerics.

Finally, we note that the trial state Eq.~\eqref{eq:psi_VBSzFM} can interpolate between the \VBSI+\zFM and \VBSII+\zFM coexistence phase regimes occurring near the corresponding pure dimer phases.
However, during this interpolation it passes through the pure \zFM state, which is formally a different phase.
Based on general arguments, we expect that the \VBSI+\zFM and \VBSII+\zFM should be in the same phase; that is, there should be a connection between the two regimes without closing the gap, and in particular with the translation symmetry broken throughout.
In App.~\ref{subapp:coexist_mps}, we will show that this is indeed possible, but we need to go beyond separable states and consider wavefunctions with entanglement across all cuts, which is achieved using an analytic MPS.

\subsection{Improved mean field states for the \texorpdfstring{$x$}{x}UUDD phase}
\label{subapp:improved_xUUDD}

Our study in App.~\ref{subapp:primitive_trial_states} simply compares trial energies of states with no variational parameters that cannot connect to each other, and in this setting the \VBS to $x$UUDD transition is first order.
A careful consideration of symmetries reveals that the true transition between either of the \VBS phases and the $x$UUDD phase should be Ising-like: both \VBS phases preserve $g_x$, $g_z$, and $T_1^2$ (or, equivalently, $g_z T_1^2$), while the $x$UUDD phase preserves $g_x$ and $g_z T_1^2$.
The two phases thus differ only by a broken $\Z_2$ symmetry, and we expect an Ising-like transition.

We can better reflect this in the mean field treatment by replacing the site-product state in Eq.~\eqref{eq:psi_xUUDD} by dimer-product states connected to the \VBS wavefunctions.
Specifically, starting from the \VBSI state, we can construct the following period-4 trial state, which is invariant under $g_x$ and $g_z T_1^2$:
\begin{widetext}
\begin{equation}
\ket{x\text{UUDD}'} = \bigotimes_n
\left[ \cos\frac{\beta}{2}\; \ket{+\hat{x}, +\hat{x}} + \sin\frac{\beta}{2}\; \ket{-\hat{x}, -\hat{x}} \right]_{4n-3, 4n-2} \otimes
\left[ \cos\frac{\beta}{2}\; \ket{-\hat{x}, -\hat{x}} + \sin\frac{\beta}{2}\; \ket{+\hat{x}, +\hat{x}} \right]_{4n-1, 4n} ~.
\label{eq:psi_xUUDD'}
\end{equation}
\end{widetext}

One observes that $\beta = \pi/2$ gives the pure \VBSI state, while $\beta = 0$ gives the $x$UUDD product state from Eq.~\eqref{eq:psi_xUUDD}.
The ground state manifold in the $x$UUDD phase is four-dimensional and is spanned by the above state with generic $\beta$ and its counterparts $T_1 \ket{x\text{UUDD}'}$, $T_1^2 \ket{x\text{UUDD}'} = g_z \ket{x\text{UUDD}'}$, and $T_1^3 \ket{x\text{UUDD}'}$.
The trial energy per site is
\begin{equation}
\epsilon_{x\text{UUDD}'} = -K_{2x} + (K_{2x} - \frac{J_x}{2}) \sin^2\beta - \frac{J_z}{2} \sin\beta ~.
\label{eq:eps_xUUDD'}
\end{equation}
For $K_{2x} < J_x/2 + |J_z|/4$, the optimal $\sin\beta = \text{sign}(J_z)$, and assuming $J_z > 0$ the state reduces to the pure \VBSI state.
For $K_{2x} > J_x/2 + |J_z|/4$, the energy is minimized by $\sin\beta = J_z/(2(2K_{2x} - J_x))$ and is given by $\epsilon_{x\text{UUDD}'} = -K_{2x} - J_z^2/(8(2 K_{2x} - J_x))$; this describes a generic $x$UUDD phase near the \VBSI phase.
The mean field transition between the two phases is continuous, which explains our observation of continuous pseudocritical behavior in VUMPS at the \VBSI to $x$UUDD transition.
However, we do not report any details of this study since it is outside our main interest.

We can also start from the \VBSII state and construct another period-4 trial state for the $x$UUDD phase that is invariant under $g_x$ and $g_z T_1^2$:
\begin{widetext}
\begin{equation}
\ket{x\text{UUDD}''} = \bigotimes_n
\left[ \cos\frac{\gamma}{2}\; \ket{+\hat{x}, -\hat{x}} + \sin\frac{\gamma}{2}\; \ket{-\hat{x}, +\hat{x}} \right]_{4n-3, 4n-2} \otimes
\left[ \cos\frac{\gamma}{2}\; \ket{-\hat{x}, +\hat{x}} + \sin\frac{\gamma}{2}\; \ket{+\hat{x}, -\hat{x}} \right]_{4n-1, 4n} ~.
\label{eq:psi_xUUDD''}
\end{equation}
\end{widetext}
Clearly, $\gamma = \pi/2$ gives the pure \VBSII state, while $\gamma = 0$ gives the primitive $x$UUDD state in Eq.~\eqref{eq:psi_xUUDD}.
The trial energy per site is
\begin{equation}
\epsilon_{x\text{UUDD}''} = -K_{2x} + (K_{2x} + \frac{J_x}{2}) \sin^2\gamma - \frac{J_z}{2} \sin\gamma ~.
\end{equation}
Comparing with Eq.~\eqref{eq:eps_xUUDD'}, we see that $\epsilon_{x\text{UUDD}''}$ has the same form as $\epsilon_{x\text{UUDD}'}$ except for the sign of the $J_x$ term.
Hence, the variational energy minimization and mean-field transition between the \VBSII and $x$UUDD$''$ state is similar to that between the \VBSI and $x$UUDD$'$ state discussed above.

We also see that for $J_x > 0$ we have $\epsilon_{x\text{UUDD}'} < \epsilon_{x\text{UUDD}''}$, and the opposite for $J_x < 0$.
As we vary $J_x$ across $J_x = 0$, since the optimal $\sin\beta = \sin\gamma = J_z/(4K_{2x}) \neq 0$, the two trial energies cross with opposite non-zero slopes; that is, we find a first-order transition between the $x$UUDD$'$ and $x$UUDD$''$ states, which are different at the transition.
One exception is the limit $K_{2x} \to \infty$ where $\beta = \gamma = 0$ and both states reduce to the site-product $x$UUDD state in Eq.~\eqref{eq:psi_xUUDD} (up to a translation).

\begin{figure}
\centering
\includegraphics[width=\columnwidth]{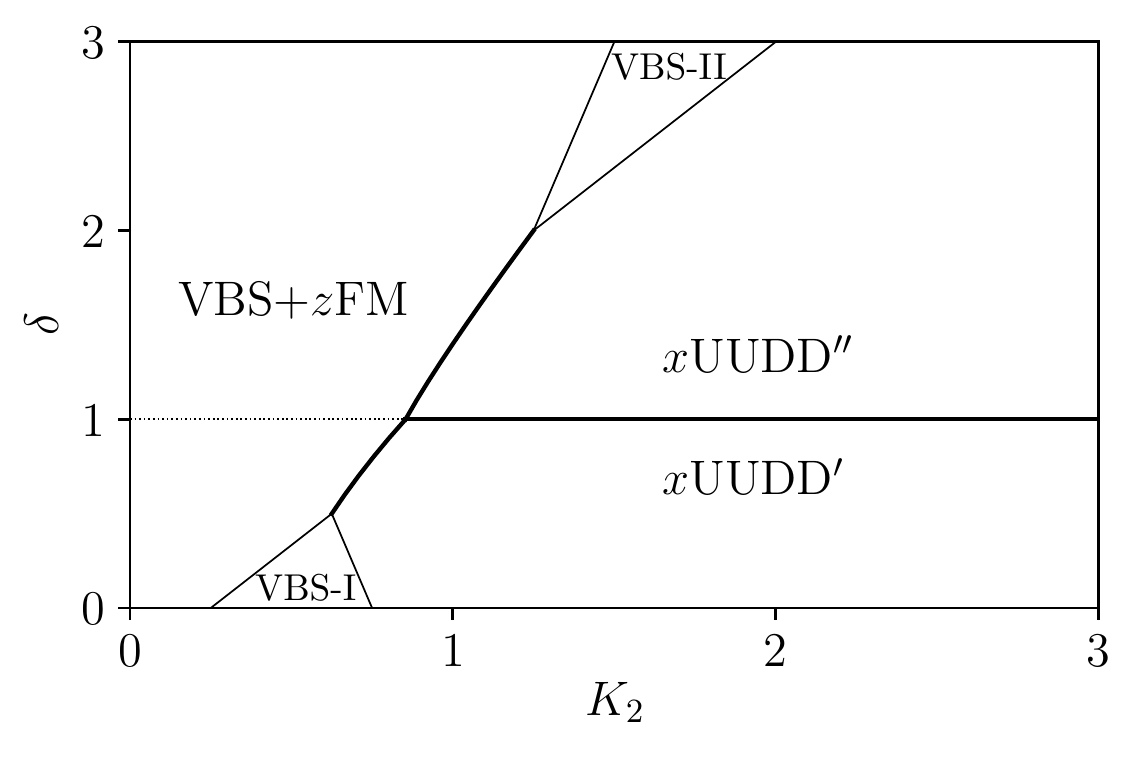}
\caption{\label{fig:mfphased2} The phase diagram of the improved mean-field trial states described in Apps.~\ref{subapp:meanfield_coexist} and \ref{subapp:improved_xUUDD} provides a somewhat more realistic picture, in particular with continuous phase transitions for all boundaries of the \VBSI and \VBSII phases.
There is an extended boundary between the coexistence and $x$UUDD phases, which is first-order, as well as a first-order transition between $x$UUDD$'$ and $x$UUDD$''$.
Along the dotted line at $\delta=1$ the \VBS+\zFM ansatz coincides with the simple product \zFM state from App.~\ref{subapp:primitive_trial_states};
however, the wider \zFM phase is not represented, as away from this special line the simple \zFM wavefunction is always energetically unfavorable.
In App.~\ref{app:meanfield_mps} we show how some of the unphysical features can be fixed using more entangled wavefunctions.
}
\end{figure}

Figure~\ref{fig:mfphased2} shows our final mean field phase diagram combing results in this section and in Sec.~\ref{subapp:meanfield_coexist}.
It includes competition between the \VBS+\zFM and $x$UUDD phases, which have incompatible symmetries and hence are separated by first-order transitions.

Regarding the first-order transition between the $x$UUDD$'$ and $x$UUDD$''$ states found in this mean field, we believe that these states are representatives of the same phase coming from different regimes, one near the \VBSI phase and the other near \VBSII.
That is, while the \VBSI and \VBSII phases are distinct phases protected by the $g_x$ and $g_z$ symmetries, $x$UUDD$'$ and $x$UUDD$''$ break $g_z$ and are not distinct phases.
One can still have a first-order transition between $x$UUDD$'$ and $x$UUDD$''$ states originating from the respective different regimes, as happens in the above mean field and is akin to a liquid-gas first-order transition. 
While this may be realized in some Hamiltonians, this does not happen in the true ground states of the model studied in this paper.
Instead we find a smooth evolution across the $\delta = 1$ line where $J_x = 0$.

As described in the main text, the $\delta = 1$ line has an additional symmetry $g_{z, \text{even}}$.
The generic $x$UUDD$'$ and $x$UUDD$''$ states considered away from this line of course do not have this symmetry but are in fact related by the action of $g_{z, \text{even}}$.
The above mean field where the two states meet discontinuously at $\delta = 1$ would correspond to spontaneously breaking the additional $\Z_2$ symmetry and hence would imply eight-fold ground state degeneracy.
In our Hamiltonian, instead it appears that the system on the $\delta = 1$ line preserves the additional $\Z_2$ symmetry, and the ground state degeneracy is four everywhere in the $x$UUDD phase.
As we show in App.~\ref{subapp:xuudd_mps}, this scenario can be also realized at the level of improved wavefunctions connected to the above $x$UUDD$'$ and $x$UUDD$''$ states but requires allowing entanglement between all sites. 

\section{Simple entangled states for phases}
\label{app:meanfield_mps}

In this appendix we add on to our mean-field treatment to address inconsistencies between the study in the main text and the mean field phase diagram obtained using only separable wavefunctions.
Specifically, in App.~\ref{subapp:coexist_mps} we write an MPS of bond dimension 2 describing the coexistence region \VBS+\zFM and matching the symmetries observed in the numerical study, which in particular can connect smoothly across the $\delta = 1$ line with the additional symmetry $g_{z,\text{even}}$.
In App.~\ref{subapp:xuudd_mps} we write another MPS of bond dimension 2 which interpolates smoothly between the improved states for the $x$UUDD phase given in Eqs.~\eqref{eq:psi_xUUDD'} and \eqref{eq:psi_xUUDD''} without a phase transition, maintaining the observed ground state degeneracy of 4 throughout.

\subsection{\texorpdfstring{$\chi = 2$}{chi = 2} MPS wavefunction for coexistence phase}
\label{subapp:coexist_mps}

In order to write a wavefunction for the \VBS and \zFM coexistence phase, we require invariance under $g_z$, $(T_1)^2$, and inversion $I$ about a bond center, and allow breaking of $g_x$ and $T_1$.
At special values of the internal parameters, our wavefunction will also be invariant under $T_1 g_{z, \text{even}}$, which is an additional symmetry present in our model at $\delta = 1$ as described in Sec.~\ref{sec:delta1}.
We use an MPS of bond dimension 2 with a two-site unit cell, having the following form:
\begin{equation}
\ket{\text{MPS}_{\VBS+\zFM}} = \sum_{\{\sigma\}} \text{Tr}[ \cdots A^{\sigma_{2m-1}} B^{\sigma_{2m}} \cdots ] \ket{\{\sigma\}} ~.
\end{equation}

The choice of the unit cell enforces $(T_1)^2$ symmetry, and we can impose invariance under $g_z$ and $I$ as follows.
A symmetry $\O$ induces on the MPS matrices an action 
$M_\O: (A^\ket{\sigma}, B^\ket{\sigma}) \mapsto (A_\O^\ket{\sigma}, B_\O^\ket{\sigma})$.
Choosing a particular representation of the projective symmetry group on the virtual indices, we can guarantee invariance of the state under $\O$ by specifying invertible matrices $X_\O$ and $Y_\O$ such that $A_\O^\ket{\sigma} = X_\O A^\ket{\sigma} Y_\O^{-1}$, $B_\O^\ket{\sigma} = Y_\O B^\ket{\sigma} X_\O^{-1}$.

Now, $g_z$ is expressed as an action on the matrices by
\begin{equation}
A_{g_z}^\ket{\sigma} = \sigma A^\ket{\sigma} ~, \quad B_{g_z}^\ket{\sigma} = \sigma B^\ket{\sigma} ~,
\end{equation}
while for bond inversion $I$,
\begin{equation}
A_I^\ket{\sigma} = (B^\ket{\sigma})^T ~, \quad B_I^\ket{\sigma} = (A^\ket{\sigma})^T ~.
\end{equation}
As we will show, the specific choice of matrices $(X_{g_z}, Y_{g_z}) = (\sigma^z, \sigma^z)$ and $(X_I, Y_I) = (\sigma^z, 1)$ allows us to connect the MPS state to the product \VBS+\zFM state \eqref{eq:psi_VBSzFM} considered earlier.
Using this choice, we find that the most general form of the MPS matrices is given by
\begin{equation}
\begin{array}{rlrl}
A^\ket{\up} &= \begin{bmatrix} a & 0 \\ 0 & b \end{bmatrix} ~, & \quad 
A^\ket{\dn} &= \begin{bmatrix} 0 & c \\ d & 0 \end{bmatrix} ~,\\ [16pt]
B^\ket{\up} &= \begin{bmatrix} a & 0 \\ 0 & -b \end{bmatrix} ~, & \quad 
B^\ket{\dn} &= \begin{bmatrix} 0 & -d \\ c & 0 \end{bmatrix} ~.
\end{array}
\end{equation}
Of the four parameters $a, b, c, d$, only three are independent, as the overall scale only affects the wavefunction normalization.

For parameters $b = d = 0$, we have $A^\ket{\up} B^\ket{\up} = \text{diag}(a^2, 0)$, $A^\ket{\dn} B^\ket{\dn} = \text{diag}(c^2, 0)$ and $A^\ket{\up} B^\ket{\dn} = A^\ket{\dn} B^\ket{\up} = 0$.
Then it is easy to see that the MPS wavefunction reduces to a form matching
the separable \VBS+\zFM wavefunction \eqref{eq:psi_VBSzFM} with dimers on the $(2m-1, 2m)$ bonds.
This state is natural near the \VBSI phase (if $c^2 \neq a^2$, it clearly breaks the $g_x$ symmetry, and approaches the \VBSI phase as $c^2 \to a^2$). 
On the other hand, for $b = c = 0$ we have $B^\ket{\up} A^\ket{\up} = \text{diag}(a^2, 0)$, $B^\ket{\dn} A^\ket{\dn} = \text{diag}(-d^2, 0)$ and $B^\ket{\up} A^\ket{\dn} = B^\ket{\dn} A^\ket{\up} = 0$.
In this case, the wavefunction reduces to a form matching
the separable \VBS+\zFM wavefunction with dimers on the $(2m, 2m+1)$ bonds.
This state is natural near the \VBSII phase.

Furthermore, we can connect the two regimes while staying within the same \VBS+\zFM phase.
For example, we can fix $a = 1$, $b = 0$, and vary between the two regimes on a path $(c, d) = \left( \gamma (1-\ell), \gamma \ell \right)$, $\ell \in [0, 1]$, with fixed $\gamma < 1$.
One can check that both $g_x$ and $T_1$ remain broken everywhere on this path.
By straightforward diagonalization of the transfer matrix one also sees that the MPS remains injective throughout the range $\ell \in [0, 1]$.
We thus conclude that the \VBSI and \VBSII orders are not distinguished in the presence of \zFM order, where $g_x$ is broken; that is, there is only one \VBS+\zFM phase.

Finally, if $c = d$ with arbitrary $a, b$, the MPS wavefunction is invariant under $S = T_1 g_{z, \text{even}}$, which is the additional symmetry present in our model on the $\delta = 1$ line.
Indeed, the action of $S$ on the above MPS induces the following action on the matrices
\begin{equation}
A_S^\ket{\sigma} = \sigma B^\ket{\sigma} ~, \quad B_S^\ket{\sigma} = A^\ket{\sigma} ~.
\end{equation}
The new matrices are gauge-equivalent to the originals under $(X_S, Y_S) = (\sigma^z, 1)$.
On the path discussed above interpolating between the \VBSI+\zFM and \VBSII+\zFM regimes, the midpoint $\ell = 1/2$ gives $c = d$ and has this symmetry.
Thus, we have also constructed candidate wavefunctions for the \VBS+\zFM coexistence phase on the $\delta = 1$ line that respect the additional symmetry present in our Hamiltonian on this line, and that appear to capture qualitative features of the true ground states of our Hamiltonian.

\subsection{\texorpdfstring{$\chi = 2$}{chi = 2} MPS for the \texorpdfstring{$x$}{x}UUDD phase}
\label{subapp:xuudd_mps}
We can write down the desired wavefunction interpolating smoothly between the separable mean field states $\ket{x\text{UUDD}'}$ of Eq.~\eqref{eq:psi_xUUDD'} and $\ket{x\text{UUDD}''}$ of Eq.~\eqref{eq:psi_xUUDD''} as a period-4 MPS with bond dimension 2 as follows:
\begin{equation}
\begin{aligned}
& \ket{\text{MPS}_{x\text{UUDD}}} = \\
& \quad \sum_{\{\sigma\}} \text{Tr}[ \cdots A^{\sigma_{4n-3}} B^{\sigma_{4n-2}} C^{\sigma_{4n-1}} D^{\sigma_{4n}} \cdots ] \ket{\{\sigma\}} ~.
\end{aligned}
\end{equation}
Here we use the $\sigma^x$ eigenbasis, and the MPS matrices are
\begin{equation*}
\begin{array}{rl}
& A^\ket{+\hat{x}} = \begin{bmatrix} r & 0 \\ 0 & s \end{bmatrix} ~, \quad
A^\ket{-\hat{x}} = \begin{bmatrix} 0 & u \\ v & 0 \end{bmatrix} ~; \\
& B^\ket{\sigma} = (A^\ket{\sigma})^T ~;\quad
C^\ket{\sigma} = A^\ket{-\sigma} ~;\quad
D^\ket{\sigma} = B^\ket{-\sigma} ~.
\end{array}
\end{equation*}
By construction, the state is invariant under inversion in the bond center between sites $4n - 3$ and $4n - 2$, and also under $g_z T_1^2$.
Furthermore, the state is invariant under $g_x$. 
As an action on the matrices, we have
\begin{equation}
M^\ket{\sigma}_{g_x} = \sigma M^\ket{\sigma} ~,
\end{equation}
and the new matrices are gauge-equivalent to the old matrices by noting that $M^\ket{\sigma}_{g_x} = \pm \sigma^z M^\ket{\sigma} \sigma^z$, where the plus sign is for $M = A, B$ and the minus sign is for $M = C, D$.
Thus, the state has the desired symmetry properties for a ground state in the generic $x$UUDD phase.

It is easy to check that when $s = 0$ and $v = 0$, the state reduces to the dimer-product state $\ket{x\text{UUDD}'}$ in Eq.~\eqref{eq:psi_xUUDD'}.
Similarly, when $s = 0$ and $u = 0$, the state reduces to $\ket{x\text{UUDD}''}$ in Eq.~\eqref{eq:psi_xUUDD''} (more precisely, the MPS yields $T_1 \ket{x\text{UUDD}''}$).

It is also easy to check that $T_1 g_{z, \text{even}}$ acts on this MPS wavefunction by interchanging $u$ and $v$.
Hence, when \mbox{$u = v$,} the state is invariant under $T_1 g_{z, \text{even}}$ and is a candidate ground state for the $x$UUDD phase along the $\delta = 1$ slice that does not break the additional $\Z_2$ symmetry present on this line.

\section{Direct phase transition at \texorpdfstring{$\delta = 1$}{delta = 1}}
\label{app:delta1}
In this appendix, we propose a field theory description which allows direct phase transition between the \VBS+\zFM coexistence phase and the $x$UUDD phase on the $\delta = 1$ line.

As we pointed out in the main text, $\delta = 1$ line admits an additional symmetry $g_{z, \text{even}} = \prod_m \sigma^z_{2m}$.
This additional symmetry plays an essential role for the direct phase transition between these two phases at $\delta = 1$.
For $\delta \neq 1$ where we do not have the $g_{z, \text{even}}$ symmetry, these two phases are either connected by a first order phase transition or by an intermediate VBS phase.

To see this, we first analyze the symmetry properties of these two phases at $\delta = 1$.
The total symmetry group at $\delta = 1$ is generated by $\{g_z, g_x, g_{z, \text{even}}, T_1, I, \TT\}$.
For our purposes here, it is enough to focus on the symmetry group generated by $\{g_z, g_x, g_{z, \text{even}}, T_1\}$.
We notice that both phases break $g_{z, \text{even}}$ and $T_1$ but preserve the combination $T_1 g_{z, \text{even}}$
\footnote{More precisely, both the \VBS+\zFM and $x$UUDD phases have ground state degeneracy equal to four.
In the \VBS+\zFM phase, all four ground states are invariant under $T_1 g_{z, \text{even}}$.
On the other hand, in the $x$UUDD phase, two of the ground states preserve $T_1 g_{z, \text{even}}$ while the other two preserve $T_1 g_{z, \text{odd}}$.}.
The \VBS+\zFM coexistence phase additionally breaks $g_x$, and the remaining symmetry group is generated by $\{g_z, T_1 g_{z, \text{even}}\}$, whereas the $x$UUDD phase breaks $g_z$, with the remaining symmetry group generated by $\{g_x, T_1 g_{z, \text{even}}\}$. 
The ground state degeneracy is four for either of these two phases.

Since the remaining symmetry groups of these two phases are not subgroups of each other, if there is a direct phase transition, this transition must be beyond the Landau--Ginzburg symmetry-breaking paradigm.
To develop a theory for this transition, we start from a background configuration that breaks $g_{z, \text{even}}$ and $T_1$, but preserves $T_1 g_{z, \text{even}}$.
For a concrete example of such a background-locking term, we can consider adding to the Hamiltonian a term $\Delta H = J_{x,\,\text{stagg}} \sum_j (-1)^j S_j^x S_{j+1}^x$.
In this background configuration, the \VBS+\zFM coexistence phase breaks $g_x$, and thus can be viewed as a ``$z$-ordered'' phase on the background.
Similarly, the $x$UUDD breaks $g_z$, and can be viewed as an ``$x$-ordered'' phase.
Hence, the phase transition can be viewed as the transition between the $z$-ordered and $x$-ordered phases on this background configuration.

Motivated by the above discussion, we can now present a hydrodynamic description for this transition.
We first define a new set of spin variables as
\begin{equation}
\begin{aligned}
& S^{\prime\, x/y}_{4n-3} = S^{x/y}_{4n-3} ~, \quad S^{\prime\, x/y}_{4n-2} = S^{x/y}_{4n-2} ~, \\
& S^{\prime\, x/y}_{4n-1} = -S^{x/y}_{4n-1} ~, \quad S^{\prime\, x/y}_{4n} = -S^{x/y}_{4n} ~; \\
& S^{\prime\, z}_j = S^z_j ~.
\end{aligned}
\label{}
\end{equation}
$T_1 g_{z, \text{even}}$ acts as a conventional translation symmetry on the new spin variables.
(For example, the specified concrete background-locking term becomes simply $\Delta H = -J_{x,\,\text{stagg}} \sum_j S_j^{\prime\, x} S_{j+1}^{\prime\, x}$.)
We then apply standard bosonization techniques on the new spins:
\begin{equation}
\begin{aligned}
& S^{\prime\, z}_j \sim \cos\phi'_j ~, \quad S^{\prime\, x}_j \sim \sin\phi'_j ~, \\
& S^{\prime\, y}_j \sim \frac{\theta'_{j+1/2} - \theta'_{j-1/2}}{\pi} ~,
\label{}
\end{aligned}
\end{equation}
where $\phi'\in [0, 2\pi)$ and $\theta' \in [0, \pi)$ are conjugate phase and phonon variables.

The symmetry transformations of $\phi'$ and $\theta'$ read
\begin{align}
g_x~: & \quad \phi' \rightarrow \pi - \phi' ~, \quad \theta' \rightarrow -\theta' ~; \\
g_z~: & \quad \phi' \rightarrow -\phi' ~, \quad \theta' \rightarrow -\theta' ~; \\
T_1 g_{z, \text{even}}~: & \quad \phi' \rightarrow \phi' ~, \quad \theta' \rightarrow \theta' + \frac{\pi}{2} ~.
\label{}
\end{align}
Thus, the symmetry-allowed scattering (i.e., cosine) terms are $\cos(2m\phi')$ and $\cos(4n\theta')$.

The action for the field theory is 
\begin{align*}
S \!=\! & \int\! \dd \tau\, \dd x \left[ \frac{\ii}{\pi} \partial_\tau \phi' \partial_x \theta' + \frac{v'}{2\pi} \! \left( \frac{1}{g'}(\partial_x \theta')^2 + g' (\partial_x \phi')^2 \right) \right] \nonumber \\
+ \!&\int\! \dd \tau\, \dd x \left[ \lambda_2 \cos(2\phi')\!+\! \lambda_4 \cos(4\phi')\!+\!\kappa_4 \cos(4\theta')\!+\cdots \right],
  \label{eq:coexistence_xUUDD_transition_field_theory}
\end{align*}
where the Luttinger parameter $g'$ and velocity $v'$ depend on microscopic details, and $\cdots$ denotes higher order scattering terms.
The scaling dimensions for the scattering terms read
\begin{align*}
  \text{dim}\left[\cos(2m\phi') \right] = \frac{m^2}{g'} ~, \quad \text{dim}\left[\cos(4n\theta') \right] = 4n^2 g' ~.
\end{align*}
In particular, when $1/2 < g' < 2$, there is only one relevant cosine operator, which is $\cos(2\phi')$.

For $\lambda_2 > 0$, $\phi'$ gets pinned at $\pi/2$ or $3\pi/2$, and thus $\lrangle{S^{\prime\,x}} \sim \lrangle{\sin\phi'} \neq 0$, which gives the $x$UUDD phase.
On the other hand, for $\lambda_2 < 0$, $\phi'$ gets pinned at $0$ or $\pi$, and thus $\lrangle{S^{\prime\, z}} \sim \lrangle{\cos\phi'} \neq 0$, which gives the \VBS+\zFM coexistence phase.
(Recall that we are working on top of a background that breaks $T_1$, which is why the ground state degeneracy is two in each case here.)
The continuous phase transition happens when $\lambda_2 = 0$, which is described by a free Luttinger liquid theory with $c = 1$ and varying critical exponents depending on $g'$.

Finally, we mention that in the absence of $T_1 g_{z, \text{even}}$, $\cos(2\theta')$ is allowed by symmetry, which becomes relevant when $g' < 2$.
It is easy to check that there are always multiple relevant or marginal operators for any $g'$. 
Thus, the above field theory loses applicability for the transition between the \VBS+\zFM and $x$UUDD phases.

\bibliography{refs.bib}

\end{document}